\documentclass[epj]{svjour}
\usepackage{amssymb}
\usepackage{graphics}
\begin{document}
\title{Systematic study of the $\rm pp \to pp \omega$ reaction}

\author{{The COSY-TOF Collaboration~ }\newline\newline
M.~Abdel-Bary\inst{3}   \and
S.~Abdel-Samad\inst{3}   \and
K.-Th.~Brinkmann\inst{1} \and
H.~Clement\inst{4}   \and
J.~Dietrich\inst{1}   \and
E.~Doroshkevich\inst{4}   \and
S.~Dshemuchadse\inst{1}   \and
K.~Ehrhardt\inst{4}   \and
A.~Erhardt\inst{4}   \and
W.~Eyrich\inst{2}   \and
A.~Filippi\inst{7}   \and
H.~Freiesleben\inst{1}   \and
M.~Fritsch\inst{2}   \and
W.~Gast\inst{3}   \and
A.~Gillitzer\inst{3}   \and
J.~Gottwald\inst{1}   \and
H.~J\"ager\inst{3}   \and
B.~Jakob\inst{1}   \and
R.~J\"akel\inst{1}   \and
L.~Karsch\inst{1}   \and
K.~Kilian\inst{3}   \and
H.~Koch\inst{8}   \and
M.~Krapp\inst{2}   \and
J.~Kre\ss\inst{4}   \and
E.~Kuhlmann\inst{1}   \and
A.~Lehmann\inst{2}   \and
S.~Marcello\inst{7}   \and
S.~Mauro\inst{8}   \and
P.~Michel\inst{5}   \and
K.~M\"oller\inst{5}   \and
H.~P.~Morsch\inst{3,} \inst{6}   \and
L.~Naumann\inst{5}   \and
N.~Paul\inst{3}   \and
C.~Pizzolotto\inst{2}   \and
Ch.~Plettner\inst{1}   \and
S.~Reimann\inst{1}   \and
M.~Richter\inst{1}   \and
J.~Ritman\inst{3}   \and
E.~Roderburg\inst{3}   \and
A.~Schamlott\inst{5}   \and
P.~Sch\"onmeier\inst{1}   \and
W.~Schroeder\inst{2}   \and
M.~Schulte-Wissermann\inst{1}   \and
T.~Sefzick\inst{3}   \and
M.~Steinke\inst{8}   \and
G.~Y.~Sun\inst{1}   \and
A.~Teufel\inst{2}   \and
W.~Ullrich\inst{1}   \and
G.~J.~Wagner\inst{4}   \and
M.~Wagner\inst{2}   \and
R.~Wenzel\inst{1}   \and
A.~Wilms\inst{8}   \and
P.~Wintz\inst{3}   \and
P.~W\"ustner\inst{3}   \and 
P.~Zupranski\inst{6}}

%

\mail{w.ullrich@physik.tu-dresden.de (W.~Ullrich)}
%
\institute{Institut f\"ur Kern- und Teilchenphysik, Technische Universit\"at Dresden, D-01062 Dresden, Germany \and
           Physikalisches Institut, Universit\"at Erlangen-N\"urnberg, D-91058 Erlangen, Germany \and
           Institut f\"ur Kernphysik, Forschungszentrum J\"ulich, D-52425 J\"ulich, Germany \and
           Physikalisches Institut, Universit\"at T\"ubingen, D-72076 T\"ubingen, Germany \and
           Institut f\"ur Kern- und Hadronenphysik, Forschungszentrum Dresden-Rossendorf, D-01314 Dresden, Germany \and
           Soltan Institute for Nuclear Studies, 05-400 Swierk/Otwock, Poland \and
           INFN Torino, 10125 Torino, Italy \and
           Institut f\"ur Experimentalphysik, Ruhr-Universit\"at Bochum, D-44780 Bochum, Germany}
\date{Received: date / Revised version: date}
%
\abstract{
  A systematic study of the production of $\omega$-mesons in
  proton-proton-collisions was carried out in a kinematically complete experiment at three excess energies
  ($\varepsilon= 92, 128, 173$ MeV). Both
  protons were detected using the large-acceptance COSY-TOF
  spectrometer at an external beam line at the Cooler Synchrotron COSY at Forschungszentrum
  J\"ulich. The total cross section, angular distributions of
  both $\omega$-mesons and protons were measured
  and presented in various reference frames such as the overall CMS, helicity and Jackson frame.
  In addition, the orientation of the $\omega$-spin and invariant mass
  spectra were determined. We observe $\omega$-production to take place
  dominantly in Ss and Sp final states at 
  $\varepsilon= 92, 128$ MeV and, additionally, in Sd at $\varepsilon= 173$ MeV.
  No obvious indication of resonant $\omega$-production via
  $N^*$-resonances was found, as proton angular distributions are almost isotropic
  and invariant mass spectra are compatible with phase space distributions.
  A dominant role of $\rm ^3P_1$ and $\rm ^1S_0$ initial partial waves
  for $\omega$-production was concluded from the orientation of the decay plane of the 
  $\omega$-meson. Although the Jackson angle distributions in the
  $\omega$p-Jackson frame are anisotropic we argue that this is not an
  indication of a resonance but rather a kinematical effect reflecting the
  anisotropy of the $\omega$ angular distribution. The helicity angle distribution
  in the $\omega$p-helicity frame shows an anisotropy which probably  
  reflects effects of the $\omega$ angular momenta in the final state; this observable may be, 
  in addition to the orientation of the $\omega$ decay plane,
  the most sensitive one to judge the validity of theoretical descriptions of 
  the production process. 
\PACS{
{13.75Cs}{Nucleon-nucleon interactions}  \and
{13.88e+}{Polarization in interactions and scattering}   \and
{14.40Cs}{Other mesons with S=C=0, mass $<$ 2.5 GeV} \and
{25.10+s}{Nuclear reactions involving few-nucleon systems}   \and
{25.40Ve}{Other reactions above meson production thresholds (energies $>$ 400 MeV)}
     } 
} 
%
\maketitle
\section{Introduction}
\label{intro}
The production of $\omega$-mesons in proton-proton collisions has come into
the focus of experimental and theoretical studies only in recent years. Early 
measurements of production cross sections employed hydrogen bubble chambers, 
a compilation of those data was published by 
Flaminio {\it et al.}\ \cite{Flaminio84} for beam momenta of
$p_{\rm beam} = 3.99\rm\; to \; 19.0\, GeV/c$. In the last 10 years several
experiments have been performed with electronic detectors, namely by the DISTO
collaboration at $p_{\rm beam}$ = 3.67 GeV/c \cite{Balestra1998},
\cite{Balestra2001} and at beam momenta below 3.2 GeV/c by Hibou {\it et al.}\ \cite{Hibou99}, Barsov {\it et al.}\ 
\cite{anke07},  and the COSY-TOF collaboration
\cite{omega2001}, \cite{omega2007}. These experimental data provide, in principle, important
information about the short range part of the nucleon-nucleon interaction which is
dominated by the exchange of the isoscalar $\omega$-meson~\cite{Machleidt}. They also furnish
elementary cross sections needed for studies of in-medium properties of
mesons. 

\begin{figure}
\resizebox{0.5\textwidth}{!}{%
  \includegraphics{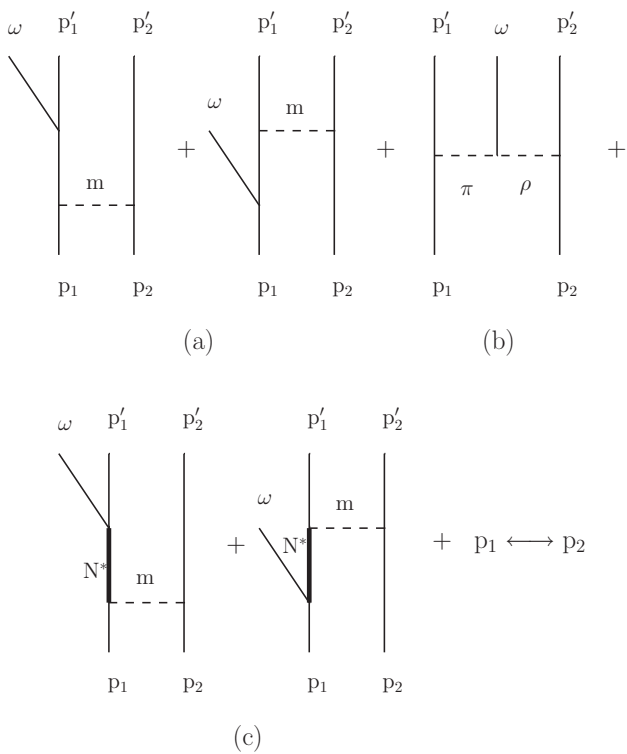}
}
\caption{Tree level diagrams for the reaction $\rm pp\to pp\omega$. Emission 
  from (a) nucleon line, (b) internal meson line, and (c)
  $N^*$-resonance. Initial and final state interaction is not indicated.} 
\label{fig:diagram}       
\end{figure}

Various theoretical models were developed to describe the experimental data. Sibirtsev \cite{Sibirtsev96} has investigated 
the $\rho, \omega$, and $\phi$-production within a one-pion exchange model. Nakayama {\it et al.}\ \cite{Nakayama98}
described the $\omega$-production within a meson exchange model taking into account nucleonic and meson-exchange currents. 
These studies were extended in \cite{Tsushima} and \cite{Nakayama07}. The $\omega$-production via nucleon resonances was studied by 
Fuchs {\it et al.}\ \cite{Fuchs03} and  Faessler {\it et al.}\ \cite{Faessler04}. 
Studies which include both resonant production and meson currents were 
performed by K\"ampfer {\it et al.}\ \cite{Kaempfer04} and Titov {\it et al.}\ \cite{Titov2002}. The diagrams considered so far in theoretical 
calculations are shown in fig.~\ref{fig:diagram}. 

These models were particularly tailored to reproduce the total cross sections and/or angular 
distributions of the $\omega$-meson. The rather limited experimental data basis does not suffice to prove one or the other 
model preferable. In addition, none of the theoretical models takes into account all 
diagrams of fig.~\ref{fig:diagram} and 
most of them neglect either initial or final state interactions or both.

In this paper we present a wealth of experimental data for the $\rm pp \to
pp\omega$ reaction obtained at three different beam momenta,
namely $p_{\rm beam} = 2.950, 3.059, 3.200$ GeV/c, which correspond to excess
energies ($\varepsilon = \sqrt s - 2 m_p - m_\omega$) 
of 92, 128, and 173 MeV, respectively. All data were taken with a special
downscaled trigger in parallel to studies of associated strangeness
procuction.  First evaluations of the data at 2.950
and 3.200 GeV/c, taken in 2000, 
focused on the angular distributions and 
total cross section for $\omega$-mesons \cite{omega2001}.
 Here we show data where the method of analysis  
was improved and extended to other observables.
The data at 3.059 GeV/c were collected in an experiment carried out with a very
large integrated luminosity in 2004 as it was devoted to the search for a
supposed pentaquark state \cite{pentaquark}. 
 Other publications of our collaboration were concerned with 
the measurement of the analysing power in the same reaction channel using a polarised 
proton beam at 3.065 GeV/c \cite{analysingpower} and a preferred orientation of the $\omega$ decay plane \cite{Ullrich2009}. 

This paper is organised as follows: The experimental procedure is detailed in sect.~2, describing the detector setup, the
principle of measurement and data analysis, Monte Carlo simulations, and a discussion of systematic uncertainties. 
In sect.~3 we present and discuss the experimental results obtained for
the total cross sections, angular distributions of the $\omega$-mesons and  
protons in the overall centre-of-mass system (CMS), the orientation of the $\omega$ decay plane, 
invariant mass spectra, and distributions of both  
helicity and Jackson angle in their respective frames. These results set important benchmarks for 
any theoretical model. The paper ends with a summary.

\section{Experimental procedure}

\subsection{Detector setup}

\begin{figure}
\resizebox{0.5\textwidth}{!}{ \includegraphics{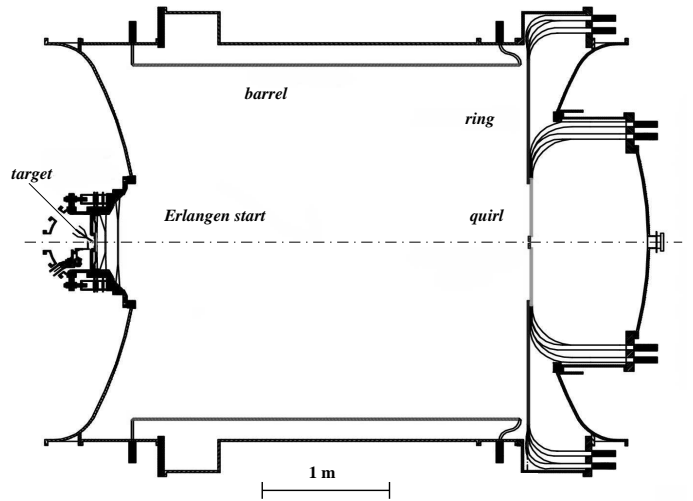} }
\caption{Sketch of the COSY-TOF spectrometer with its main components; see text for details.}
\label{fig:detector}  
\end{figure}

The experiments were carried out with the time-of-flight spectrometer COSY-TOF installed at an external beam line 
of the COoler SYnchrotron COSY at Forschungszentrum J\"ulich.
A sketch of the detector is shown in fig.~\ref{fig:detector}. 

\begin{figure*}
\resizebox{1.\textwidth}{!}{ \includegraphics{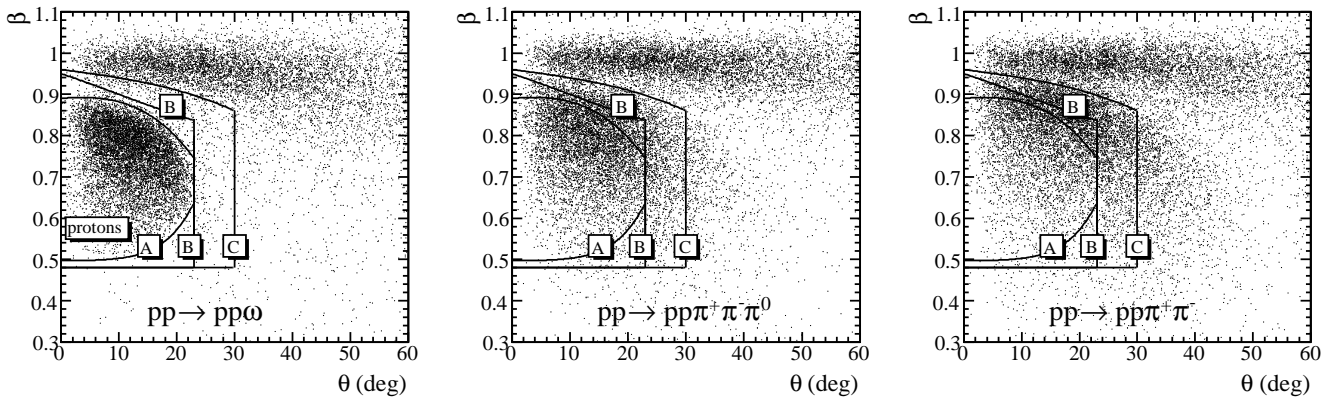} }
\caption{Monte Carlo data for pp$\to$pp$\omega$ ($\omega\to\pi^+\pi^-\pi^0$),
      pp$\to$pp$\pi^+\pi^-\pi^0$ and pp$\to$pp$\pi^+\pi^-$ at $\varepsilon$ = 128 MeV. 
      The experimental resolution of the detector has been accounted for. 
      Three regions (A, B, C) for the separation of protons and pions are indicated. See text for details.}
\label{fig:boxomega}  
\end{figure*}

The extracted proton beam (spill length $\approx$ 5 min, several $10^6$
protons/s) is directed onto a liquid hydrogen target of 6 mm diameter and 4 mm length \cite{target} 
through annular veto detectors of different diameter positioned at 260, 51 and 3.5 cm  upstream of the target.
The emerging particles traverse a layered system of time-of-flight start and tracking detectors 
(called Erlangen start \cite{Erlangenstart}).
After a flight path of $\approx\negthinspace$ 3 m through the evacuated tank (0.2 Pa) all charged particles 
are detected in the highly granular stop components of the spectrometer. They consist of two three-layered 
forward hodoscopes (quirl, ring) which are built according to the layout described in \cite{forewardhodo}, 
and a barrel hodoscope \cite{barrel}, all utilising BC412 scintillating material. 

From the combined measurement of time and position the velocity vectors of all charged particles
are determined with a time-of-flight resolution of better than $\sigma_{\rm TOF}$ = 300 ps 
and an angular track-resolution of better than $\sigma_{\sphericalangle}=0.3^\circ$. 
Due to the low mass area density of the time-of-flight start and tracking detectors, 
the influence of small angle scattering and energy loss is almost negligible for
particles with $\beta>0.5$. Only particles with these velocities are produced in the reaction under study.

Unlike magnetic spectrometers, which often provide particle identification at the cost 
of limited acceptance, the COSY-TOF spectrometer stands out for its large geometrical acceptance 
($1^\circ<\theta_{\rm lab}<60^\circ,\, 0^\circ<\phi<360^\circ$) 
and an efficiency $>95\%$ for the detection of a charged particle. 
This allows the almost unambiguous and simultaneous identification of different reaction channels 
(e.g., pp$\to$ pp, d$\pi^+$, pp$\pi^+\pi^-$, pp$\omega$, pK$^+\rm\Lambda$, pK$^+\rm\Sigma^0$, pK$^0\rm\Sigma^+$) by 
exploiting the time of flight of all emerging charged particles as well as the event's topology.

\subsection{Principle of measurement and data analysis}

The reaction $\rm pp\to pp\omega$ with the main decay channel of the
$\omega$-meson ($\omega\to\pi^+\pi^-\pi^0, \mathcal{BR}\approx89.2$\%
\cite{pdg}) is preselected during data taking via a trigger set on four
charged particles detected in the stop components (quirl, ring, barrel). The
outgoing protons and the charged pions are, due to their large mass
difference, clearly separated in a velocity versus polar angle representation
as can be seen from the left frame of fig.~\ref{fig:boxomega}; cf. also
\cite{omega2007}. 

This kinematical separation serves to identify protons as they are restricted to a region 
with $\theta < 23^\circ$ and moderate velocities, while the pions mainly cluster at $\beta\approx1$ over the 
full angular range. There is only a small number of events (about 10 to 15\%
depending on excess energy) where pions are found in the proton region.
For the analysis we require exactly two particles within one of the indicated proton regions A, B, or C, 
shown in the left frame of fig.~\ref{fig:boxomega} (the effect of these different regions will be discussed below), 
and exactly two particles outside, which are, in turn, assumed to be pions. In addition, all tracks 
must fall into the region of optimum geometrical acceptance ($3^\circ<\theta<60^\circ$). Proton and pion masses 
are then assigned accordingly and the four-momenta of protons are calculated
using the measured velocity vectors. Monte Carlo
simulations showed that the assignment of protons and pions to the
pp$\to$pp$\omega$ reaction channel based on these criteria is correct for
99.2$\%$ of all events. We would like to point out that the geometrical
acceptance for 
$\rm pp \to pp\omega \to pp \pi^+\pi^-\pi^0$ reactions, 
that means the fraction of events
with all 4 charged particles in the geometrical covered region,
is 65\% at
the lower excess energy $\varepsilon = 92\mathrm{MeV}$ and increases to 68\%
at $\varepsilon = 173\mathrm{MeV}$. This applies for phase space and isotropic
distributed events. 

The missing mass distribution calculated from the momenta of the identified protons exhibits a peak at the 
$\omega$-mass. However, this spectrum contains a large contribution of resonant two pion production 
via pp$\to$ pp$\rho,\, \rho\to\pi^+\pi^-$ 
and non-resonant production of two pions ($\rm pp\to pp\pi^+\pi^-$) as well as three pions (pp$\to$pp$\pi^+\pi^-\pi^0$), 
all of them constituting the major components of an unavoidable physical 
background. In order to accentuate this point, we show in the middle (right hand) frame of 
fig.~\ref{fig:boxomega} Monte-Carlo simulated data for 
non-resonant $\pi^+\pi^-\pi^0$ ($\pi^+\pi^-$) production.
It is obvious that in these two cases the kinematical separation of protons 
and pions is less certain. 

It is possible to reduce the two-pion background by exploiting the topology of 
the $\omega$-decay into three pions. Here, the plane defined by the two
charged pions does, in general, not contain the 
pp-missing momentum vector ($=\omega$-momentum vector) due to the momentum of the 
undetected $\pi^0$. By investigating Monte Carlo data it was found that requiring an acoplanarity angle 
$\alpha = \angle ((\overrightarrow{p}_{\pi_1}\times\overrightarrow{p}_{\pi_2}),(\overrightarrow{p}_{\pi_1}\times\overrightarrow{p}_{\omega}))>5^\circ$
rejects 90\% of the two-pion background while only 17\% of the $\omega$-events are removed, thus improving
the signal-to-background ratio. We would like to stress the robustness of the method as the selection of different 
acoplanarity angles result in the same yield of the $\omega$-signal after acceptance correction, 
within an uncertainty of 2\% \cite{omega2007}. Throughout the analysis an acoplanarity angle of $\alpha \geq 5^\circ$ is required.
Finally, in the analysis only those events are taken into account for which the sum of the proton momenta points into the 
backward hemisphere of the overall CMS in order to improve on the momentum resolution. Because of the identical particles 
in the entrance channel all distributions 
of final state particles in the overall CMS will be symmetric so that a reduction of information does not ensue.

\begin{figure}
\resizebox{0.5\textwidth}{!}{ \includegraphics{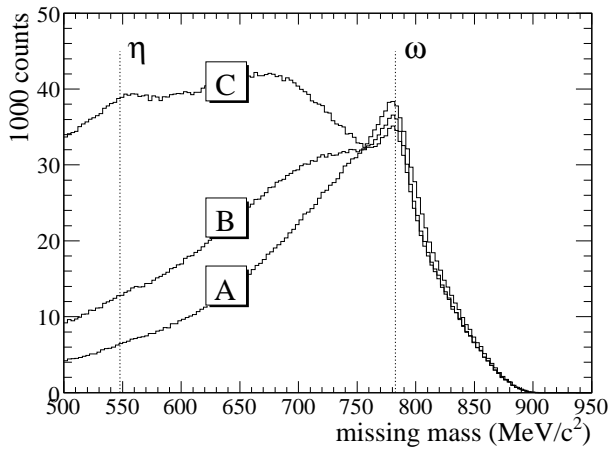} }
\caption{Proton-proton missing mass spectra at $\varepsilon= 128\,\mathrm{MeV}$ 
      obtained when applying the different regions for proton assignment shown in fig.~\ref{fig:boxomega}.}
\label{fig:mmcompare}  
\end{figure}

In the present paper our former study \cite{omega2007} is buttressed 
by scrutinising the influence of the proton region chosen on the yield determined for the $\omega$-signal. 
As already mentioned, three proton regions are considered, region A with rather narrow boundaries, 
a wider region B (formerly used in \cite{omega2007}) 
and an even larger one, region C. As shown in fig.~\ref{fig:mmcompare}, a distinctive peak at 
the $\omega$-pole mass ($m_{\omega}=$782.7 MeV/c$^2$, \cite{pdg} )
can be seen in all cases above the multi-pion background, the shape of which for missing masses smaller than the 
$\omega$-mass depends dramatically on the region chosen for the separation of protons and pions, while the shape 
above the $\omega$-mass is barely different.

Region A, which sharply restricts the protons in angle and velocity, results in 
a missing mass spectrum where the shape of the background can hardly be estimated in the region of 
the $\omega$-mass. The missing mass spectrum resulting from region B
exhibits an $\omega$-signal which resides on a smooth, apparently convex background which extends to missing 
masses smaller than the $\omega$-mass. When further enlarging the proton region to region C the background  
dominates the spectrum and its shape below the $\omega$-signal tends 
to change from a convex to an almost linear shape. The advantage of using region B \cite{omega2007} lies in the best 
signal-to-background ratio, while in case of region C the $\omega$-signal becomes less susceptible to the method 
used for its extraction (see below). This results in a smaller uncertainty for the extracted counts, although
about 20$\%$ of the signal is lost. Thus, in contrast to our earlier
analyses \cite{omega2007} we now use the wide region C which
is adjusted with increasing excess energy in order to accommodate the protons 
which become less confined in angle and velocity.
It is interesting to note that a signal of the $\eta$-meson appears at $550\,\rm MeV/c^2$,
however, the excess energy of about 350 MeV in this channel causes a large width of this signal 
greatly complicating a detailed analysis of the data. 

\begin{figure}
\resizebox{0.5\textwidth}{!}{ \includegraphics{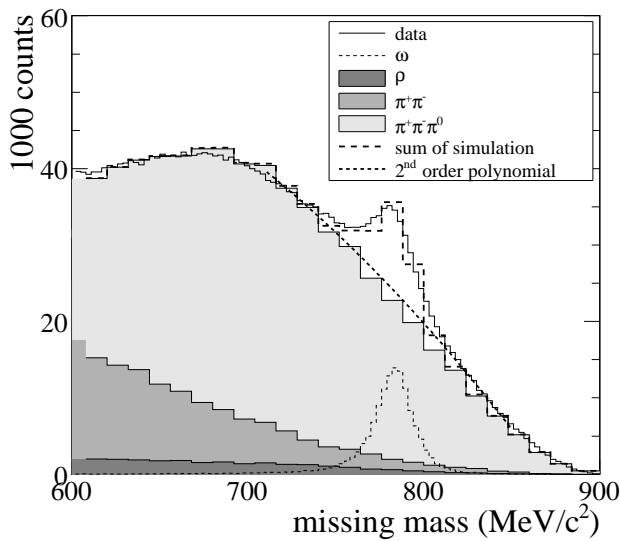} }
\caption{Proton-proton missing mass spectra at $\varepsilon= $ 128 MeV 
      with adjusted Monte Carlo distributions obtained for resonant pp$\to$pp$\rho,\, \rho\to\pi^+\pi^-$ production 
      and non-resonant two- and three-pion production. 
      The sum describes the background below the $\omega$-signal very well. (See text
      for details.)}
\label{fig:mcfit}  
\end{figure}

In order to quantitatively determine the yield of the $\omega$-signal above
background we follow two different approaches,
briefly described in \cite{Ullrich2009}. One method is a simultaneous least square fitting 
of a Voigt function (convolution of a Gaussian and a Breit-Wigner function) for the $\omega$-signal
and a second-order polynomial for the background below the signal. The width of the Breit-Wigner distribution 
is fixed to the natural width of the $\omega$-meson ($\Gamma$ = 8.49 MeV, \cite{pdg}), 
while all other parameters are allowed to vary freely. The alternative method is based on the simulation 
of the main background channels ($\rm pp\to pp\rho$, $\rho\to\pi^+\pi^-$, $\rm pp\to pp\pi^+\pi^-$, 
$\rm pp\to pp\pi^+\pi^-\pi^0$) where the simulated data are analysed in the
very same way as the experimental data. The yield of various background
contributions obtained are adjusted in order to reproduce the experimental
background. It is found that the 3$\pi$-background dominates over the
2$\pi$-background (which was effectively reduced by the acoplanarity requirement); the
non resonant 2$\pi$-contribution can be exchanged for the resonant one from 
$\rho$-decay as the $\rho$-meson has a large width. The result of both
procedures is shown in fig.~\ref{fig:mcfit}  for an excess energy of 128
MeV. The background described by the ``pion-cocktail'' agrees very well with
the one determined by the fitting procedure.
The yield determined for the $\omega$-signal is found to agree  
within a few percent for the two methods; the present results are fully
consistent with but more precise than our earlier ones.
Although this experiment does not aim at a measurement of the $\omega$ mass 
we like to quote the parameters determined for the Voigt function at
the three excess energies of $\varepsilon=92, 128, 173 \,\mathrm{MeV}$, namely $m_\omega=780.2, 782.4,
779.5\,\mathrm{MeV/c^2}$ with a systematic uncertainty of 0.3\%
and a standard deviation of the Gaussian resolution function of
$\sigma= 7.6\pm1.1, 10.3\pm0.5, 11.3\pm0.6 \,\mathrm{MeV/c^2}$.
As usual, the best momentum and, hence, mass resolution for time-of-flight experiments
is achieved for the slowest particles, i.e. protons at the smallest excess
energy. 

The procedure of yield determination must be carried
through for each interval of the differential observables, e.g.\ for different intervals 
in $\cos\theta^*_\omega$ to obtain the $\omega$ angular distribution
in the overall CMS. 

The absolute normalisation is accomplished by evaluating elastic scattering, which is always measured 
simultaneously at COSY-TOF. The angular distribution determined is
compared to literature data \cite{AltmeierEDDA}, whereupon the normalisation factor directly yields the 
integrated luminosity. The uncertainty of this procedure (5\%) is in equal parts 
due to the intrinsic uncertainty of our measurement and the error of the literature data. 

\subsection{Monte Carlo simulations}

During the analysis of the experimental data each step of the analysis was checked against results obtained by Monte Carlo 
simulations, thereby giving insight into the detector performance and, in particular, its acceptance.
The programme package \cite{Brand,Zielinsky}
models the whole COSY-TOF-detector in great detail (geometry of all sub-detectors,
detector housings and wrappings, cabling, supporting 
structures etc.). The particles are generated according to equal population of phase space \cite{Genbod} 
and propagated in space and time accounting for small angle scattering,
nuclear reactions, $\delta$-electrons and energy loss in all detector components; 
the energy loss value is used to generate a charge or light output signal. 
The AD-conversion takes into account light attenuation in the hodoscopes, 
efficiency and gain of the individual photomultipliers.
The time and charge resolution of the individual 
elements of different sub-detectors is also accounted for. 
A properly chosen amount of noise is added to each signal \cite{Karsch}.
In addition, various decay anisotropies, e.g. that 
of the $\omega$-decay ($J^P=1^-$) can be included optionally \cite{MSW}.
The Monte Carlo simulations also account for the anisotropic
$\cos\theta^*_\omega$ angular distributions 
as experimentally observed (see section 3.2.1) at the three excess energies. 
These Monte-Carlo data are subjected to the very same analysis 
routines as the experimental data in order to treat both on an equal footing,
thus, for all observables the detector acceptance can be
determined. It is found that the acceptance is a smooth function for all
differential observables as can be inferred from the figs.
\ref{fig:cosw} to \ref{fig:finjacksonwp} shown below.
The numerical value of the integral acceptance, however, depends largely on the
size of region C chosen for proton identification and turns out to be slightly
different for the three excess energies studied (see table~\ref{tab:totalx}).

\subsection{Systematic uncertainties}

Three major sources of systematic errors were identified for the present analysis. An obvious one is the uncertainty of 
the luminosity determination of 5\% (see above). 
The second source of systematic uncertainties stems from the determination of the yield of the $\omega$-signal residing 
on a sizable background.
The acceptance correction itself is the third source of systematic uncertainties. 
The Monte Carlo simulation should model, besides other observables,  
the relation $\beta$ vs $\theta_{\rm lab}$ as perfectly as possible since it is
used for particle identification. Although the majority of pions cluster at
$\beta\approx1$ some happen to fall into the $\beta$ vs $\theta_{\rm lab}$-region
chosen for proton assignment. This fraction is fairly sensitive to the correct
simulation of the time resolution for pion tracks. Its influence can be
studied by varying the $\beta$-boundary of the proton region.  Thereby the
systematic uncertainty of the acceptance correction was determined to amount
to 4\% at $\varepsilon=92\,\rm MeV$ and $\varepsilon=128\,\rm MeV$ and 6\% at
$\varepsilon=173\,\rm MeV$. 

In the case of the total cross section these three systematic uncertainties were
added quadratically.

In case of differential cross sections the systematic uncertainties for the
determination of the yield of the $\omega$-signal and the acceptance
correction were investigated in detail for each bin of the observable under
consideration. 
A spread of
uncertainties was quantified, firstly, by a systematic variation of the range
of the missing mass included when fitting the background below and above the
$\omega$-signal and, secondly, by a comparison of the results obtained by the
two different fitting methods. The spread was found to be between 5 and 20\% and
was added to the statistical error. The sum is indicated as a common error bar for
each interval of the observable under consideration. 
The systematic error due to the luminosity determination contributes an overall uncertainty
which is not shown. 

\section{Discussion of results}

\subsection{Total cross section}

The number of $\omega$-events extracted from the proton-proton missing mass spectrum at different 
excess energies are used to calculate the total cross sections listed in
table~\ref{tab:totalx}.
The values at $\varepsilon$ = 92 MeV and 173 MeV agree with those published earlier
 \cite{omega2001}, \cite{omega2007}, however with improved accuracy. 

\begin{table}
  \caption{Total cross section for the reaction $\rm pp \to pp\omega$ at three excess energies; 
    $N_\omega$: number of $\omega$-events; acc: detector acceptance as determined from 
    Monte Carlo simulations; see text for details.}
    \label{tab:totalx}
      \begin{tabular}{@{}rrrrr}
      \hline\noalign{\smallskip}
      $\varepsilon$ $(\mathrm{MeV})$ & $N_{\omega}$ & acc & $\int{L{\mathrm d}t}$ $({\mu b^{-1}})$ & $\sigma_{\rm tot}$ $(\mu b)$\\
      \noalign{\smallskip}\hline\noalign{\smallskip}
      $92$    & 2052    &  0.156  & 1449    &  9.1  $\pm$ 0.6 $\pm$ 1.0 \\  
      $128$   & 116315 &  0.168  & 53500    &  12.9 $\pm$ 0.2 $\pm$ 1.1 \\  
      $173$   & 3558    &  0.155  & 795     &  28.9 $\pm$ 1.9 $\pm$ 3.1 \\ 
      \noalign{\smallskip}\hline
    \end{tabular}
\end{table}

\begin{figure}
\resizebox{0.5\textwidth}{!}{ \includegraphics{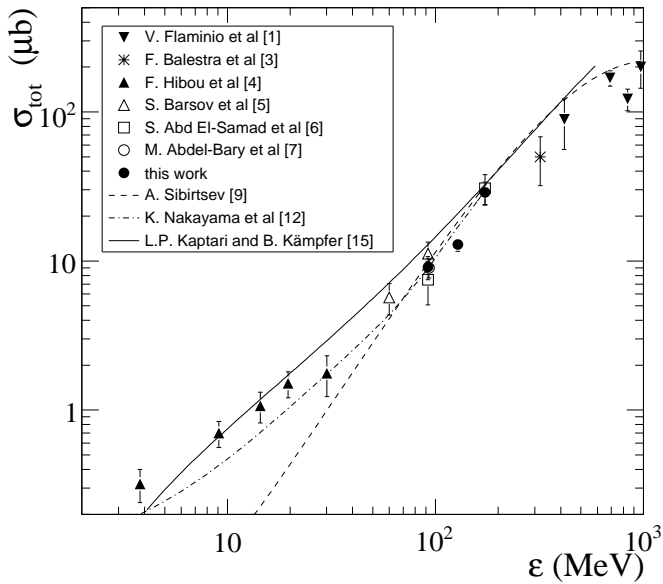} }
\caption{Total cross section as a function of excess energy measured for the reaction $\rm pp \to pp\omega$.}
\label{fig:totalx}  
\end{figure}

Fig.~\ref{fig:totalx} shows all published data for this reaction; the present data follow the general trend. The 
results of theoretical models more or less follow the data, but the total cross section 
is not a sensitive observable that proves one or the other model preferable - additional observables are needed. 
As COSY-TOF almost completely covers the phase
space of the reaction many differential cross section can be determined offering a variety of other
physical quantities setting benchmarks for theoretical models. Hence we turn to differential distributions. 

\subsection{Angular distributions of the $\omega$-mesons and protons in the overall CMS}

\subsubsection {$\omega$-mesons}

\begin{figure*}
\resizebox{1.\textwidth}{!}{ \includegraphics{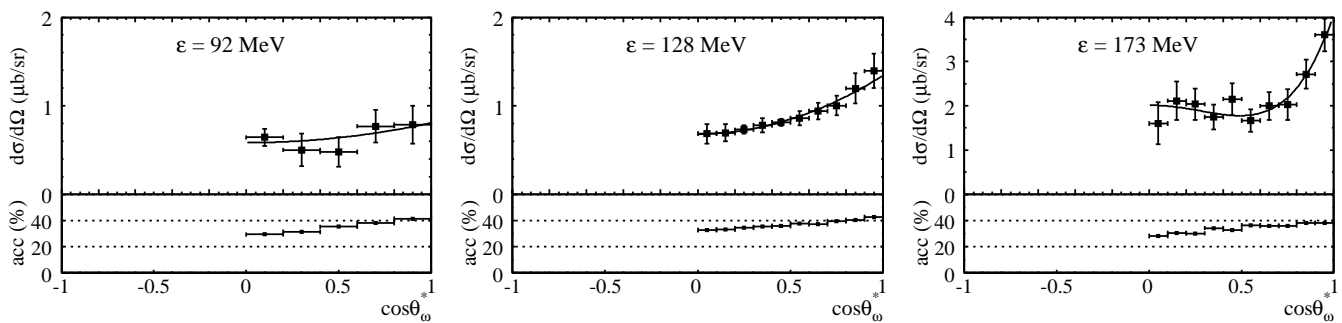} }
\caption{Angular distribution of the $\omega$-meson in the overall CMS (upper part in each frame).
      The lower part shows the acceptance of the COSY-TOF spectrometer as determined from Monte Carlo simulations.}
\label{fig:cosw}  
\end{figure*}

The $\omega$ angular distributions in the overall CMS are shown in fig.~\ref{fig:cosw} for the three excess energies, the acceptance
is shown in the lower part of each frame. 
It is a smooth function of $\cos\theta^*_{\omega}$ which drops with increasing angle only slightly from 
$\approx$40 to $\approx$30\%. As pointed out above, the 
data analysis requires two protons with an added momentum pointing into the backward hemisphere, hence we can show only 
the angular distribution for the forward hemisphere. 
A transition is observed from a slightly anisotropic to a pronounced anisotropic
angular distribution.

The standard parametrisation of an angular distribution in the overall CMS in terms of Legendre polynomials reduces to 
\begin{equation}
  {{\rm d}\sigma\over{\rm d}\Omega}=\sum _{ L=0}^{L_{max}} a_{2L}\cdot P_{2L},\quad{L=0,1,2,\dots}
  \label{eq:legendrefit}
\end{equation}
since only even Legendre polynomials $P_{2L}$ need to be taken into account in the case of
identical particles in the entrance channel. This equation can be rearranged to
\begin{equation}
  {{\rm d}\sigma\over{\rm d}\Omega}={\sigma_{\rm tot}\over {4\pi}}\cdot(1 +
  \sum _{L=1}^{L_{\rm max}} {a'_{2L}}\cdot P_{2L}),\quad{L=1,2,\dots}
  \label{eq:legendrefitsigma}
\end{equation}
where ${\sigma_{\rm tot}/{4\pi}} = a_0$ and $a'_{2L}=a_{2L}/a_0$. 
We prefer the use of eq.(\ref{eq:legendrefitsigma}) over eq.(\ref{eq:legendrefit}) as the former yields the total cross
section $\sigma_{\rm tot}$ without further calculation being necessary 
and the significance of the contributions of higher partial waves in the exit 
channel is better judged via 
the ratio $a'_{2L}$ rather than $a_{2L}$. The coefficients obtained are listed in table~\ref{tab:cosw}. 

\begin{table}
  \caption{Coefficients of Legendre polynomials determined by fitting eq.(\ref{eq:legendrefitsigma})
    to the $\omega$ angular distributions.}
    \label{tab:cosw}
      \begin{tabular}{@{}rrrr}
      \hline\noalign{\smallskip}
      $\varepsilon\;(\mathrm{MeV})$ & $\sigma_{\rm tot}\;(\mathrm{\mu b})$  & $a'_2$ &  $a'_4$ \\
      \noalign{\smallskip}\hline\noalign{\smallskip}
 $92$ & $ 9.3 \pm 1.1$ & $0.23 \pm 0.26$ & -- \\
$128$ & $12.7 \pm 0.5$ & $0.49 \pm 0.10$ & -- \\
$173$ & $30.5 \pm 1.8$ & $0.46 \pm 0.15$ & $0.42 \pm 0.18$\\
      \noalign{\smallskip}\hline
    \end{tabular}
\end{table}

While it is sufficient to use $L_{\rm max}=1$ for the two lower excess energies, as can be deduced from table~\ref{tab:cosw}
the inclusion of $L_{\rm max}=2$ improves the description of the angular
distribution for $\varepsilon$=173 MeV although the uncertainty of $a'_4$ is
rather large. The values of $\sigma_{\rm tot}$ are consistent, within the uncertainties,
with those listed in table~\ref{tab:totalx}. We like to point out that the
deduced values of  $L_{\rm max}$ are fully consistent with kinematic
considerations: the maximal orbital angular momentum for the (pp)-$\omega$
system can be estimated as $\eta = p^*$/$m_\pi$ where $p^*$ is the centre of
mass momentum in the final state, 1/(2$m_\pi$) the strong interaction radius
and $m_\pi$ the pion mass \cite{Faessler04}.
For the three excess energies $\eta$ varies from 2.4 to 3.3 $\hbar$. However, the angular momentum barrier 
truncates this maximum orbital angular momentum to an effective one of 
$L=1 \hbar$ for $\varepsilon$ = 92 and 128 MeV, while $L=2\hbar$ is only probable at $\varepsilon$ = 173 MeV. 

These results can be looked at from a different point of view. Meson
production in pp reactions is usually described by defining the
non-relativistic momenta $\vec p$ (relative momentum of the two protons) and
$\vec q$ (momentum of the $\omega$-meson with respect to the centre of mass of
the two protons) and the associated orbital angular momentum quantum numbers
$L$ and $l_{\omega}$, respectively. Then the spectroscopic notation
$(^{2S+1}L_J)_i$ holds for the partial waves in the initial state and
$(^{2S+1}L_J)_f$ $l_{\omega}$ for the final state which in short will be
designated as type "$Ll_{\omega}$" later on; $S$, $L$, $J$ refer to the
quantum numbers of spin, orbital angular momentum, and total angular
momentum of the pp-system  in either initial (i) or final (f) state (see
e.g. Hanhart \cite{hanhart}). We list
in table~\ref{tab:pw} all partial waves of the entrance channel (antisymmetric
pp-wave functions) which can produce (accounting for conservation of both parity and angular
momentum) final states with at most two units of orbital angular
momentum as suggested by the angular distribution of the $\omega$-meson. 

\begin{table}
    \caption{List of all partial waves in the initial state which can produce final states
      with two units of angular momentum at the most. The notation is defined in the text.}
    \label{tab:pw}
    \begin{tabular}{@{}lll}
    \hline\noalign{\smallskip}
     type        & $(^{2S+1}L_{J})_i\rightarrow{}(^{2S+1}L_J)_f l_\omega$  & amplitude \\
    \noalign{\smallskip}\hline\noalign{\smallskip}
     Ss          & $^3\mathrm{P}_1\rightarrow{}^1\mathrm{S}_0\,\mathrm{s}$       & $f_1$\\ \noalign{\smallskip}\hline\noalign{\smallskip}
     Sp          & $^1\mathrm{S}_0\rightarrow{}^1\mathrm{S}_0\,\mathrm{p}$       & $f_2$\\  
                 & $^1\mathrm{D}_2\rightarrow{}^1\mathrm{S}_0\,\mathrm{p}$       & $f_3$\\ \noalign{\smallskip}\hline\noalign{\smallskip}
     Ps          & $^1\mathrm{S}_0\rightarrow{}^3\mathrm{P}_1\,\mathrm{s}$       & $f_4$\\  
                 & $^1\mathrm{D}_2\rightarrow{}^3\mathrm{P}_1\,\mathrm{s}$       & $f_5$\\ 
                 & $^1\mathrm{D}_2\rightarrow{}^3\mathrm{P}_2\,\mathrm{s}$       & $f_6$\\ \noalign{\smallskip}\hline\noalign{\smallskip}
     Sd          & $^3\mathrm{P}_2\rightarrow{}^1\mathrm{S}_0\,\mathrm{d}$       & $f_7$\\  
                 & $^3\mathrm{F}_2\rightarrow{}^1\mathrm{S}_0\,\mathrm{d}$       & $f_8$\\ 
                 & $^3\mathrm{F}_3\rightarrow{}^1\mathrm{S}_0\,\mathrm{d}$       & $f_9$\\ \noalign{\smallskip}\hline\noalign{\smallskip}
     Pp          & $^3\mathrm{P}_0\rightarrow{}^3\mathrm{P}_{0,1,2}\,\mathrm{p}$ & ...\\ 
                 & $^3\mathrm{P}_1\rightarrow{}^3\mathrm{P}_{0,1,2}\,\mathrm{p}$ & ...\\ 
                 & $^3\mathrm{P}_2\rightarrow{}^3\mathrm{P}_{0,1,2}\,\mathrm{p}$ & ...\\ 
                 & $^3\mathrm{F}_2\rightarrow{}^3\mathrm{P}_{0,1,2}\,\mathrm{p}$ & ...\\ 
                 & $^3\mathrm{F}_3\rightarrow{}^3\mathrm{P}_{1,2}\,\mathrm{p}$   & ..\\ 
                 & $^3\mathrm{F}_4\rightarrow{}^3\mathrm{P}_2\,\mathrm{p}$       & .\\ \noalign{\smallskip}\hline\noalign{\smallskip}
     Ds          & $^3\mathrm{P}_2\rightarrow{}^1\mathrm{D}_2\,\mathrm{s}$       & $f_{25}$\\ 
                 & $^3\mathrm{F}_2\rightarrow{}^1\mathrm{D}_2\,\mathrm{s}$       & $f_{26}$\\ 
                 & $^3\mathrm{F}_3\rightarrow{}^1\mathrm{D}_2\,\mathrm{s}$       & $f_{27}$\\ 
    \noalign{\smallskip}\hline

    \end{tabular}
\end{table}

A comparison with the value of $L_{\rm max}$ deduced from the angular distribution shows
that at $\varepsilon$=92 and 128 MeV final states of type Ss, Sp and Ps (amplitudes $\rm f_1,f_2, .., f_6$) contribute, 
while at $\varepsilon$=173 MeV the
final states of type Sd, Pp, and Ds (amplitudes $\rm f_7, ..., f_{27}$) come into play. It is important to realise that from a 
particular initial state two or more final states can be populated (cf. amplitudes $f_2$ and $f_4$ or $f_3$, $f_5$, and $f_6$).
Which of these amplitudes are present in the reaction and contribute with which weight 
depends on the transition matrix elements; they can be determined by a full model calculation. 
However, we conjecture that, due to the available excess energy, the two final state protons are most 
probably in the $\rm ^1S_0$ state. This would reduce the mainly contributing final states 
to those of type Ss, Sp, and Sd. It cannot be excluded that final states of type Ps are present
as they originate from the same initial state as Sp type states, but probably with only little weight.

\subsubsection{Protons}

\begin{figure*}
\resizebox{1.\textwidth}{!}{ \includegraphics{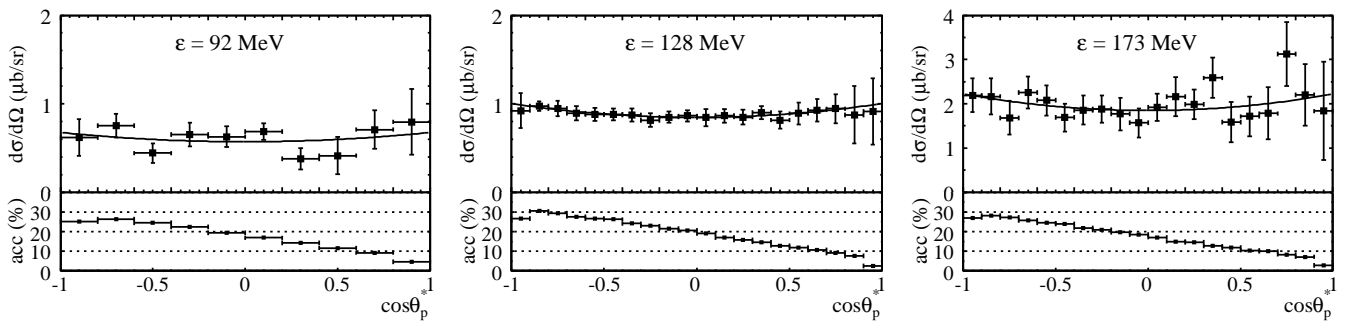} }
\caption{Angular distribution of the protons in the overall CMS (upper part in each frame). The lower part 
      depicts the acceptance of the COSY-TOF spectrometer as determined from Monte Carlo simulations.}
\label{fig:cosp}  
\end{figure*}

In fig.~\ref{fig:cosp} we show the angular distributions of protons in the
overall CMS for the three excess energies. The acceptance (lower part of each
frame) is a smooth function of $\cos\theta^*_{\rm p}$ with an almost constant
slope. The decrease towards $\cos\theta^*_{\rm p}=1$ is due to the analysis
requirement of the sum of the proton momenta pointing into the backward
direction. The angular distributions are, within uncertainties, symmetric with respect to
$\cos\theta^* = 0$, as they should be in the case of identical particles in
the entrance channel. This is a strong indication that the acceptance corrections were performed
properly.  

Using even Legendre polynomials ($\rm P_0$ and $\rm P_2)$ according to eq. (\ref{eq:legendrefitsigma})
the least square fitting resulted in the values of table~\ref{tab:cosp}. The distributions are almost isotropic; 
little room is left for an $L=1$ partial wave. As a consequence, the final state 
pp-system is most probably in the $\rm ^1S_0$-state and less probably in the $\rm ^3P$-state; this conclusion is consistent with that 
drawn from the $\omega$ angular distribution. 

\begin{table}
  \caption{Coefficients of Legendre polynomials determined by fitting eq. (\ref{eq:legendrefitsigma})
    to the proton angular distributions.} 
  \label{tab:cosp}
  \begin{tabular}{@{}rrr}      
      \hline\noalign{\smallskip}                             
      $\varepsilon\;(\mathrm{MeV})$ & $\sigma_{\rm tot}\;(\mathrm{\mu b})$  & $a'_2$ \\
      \noalign{\smallskip}\hline\noalign{\smallskip}                             
 $92$ & $ 8.5 \pm 0.7$ & $0.11 \pm 0.22$ \\
$128$ & $12.7 \pm 0.3$ & $0.12 \pm 0.06$ \\
$173$ & $27.8 \pm 1.3$ & $0.13 \pm 0.11$ \\
      \noalign{\smallskip}\hline
    \end{tabular}                 
\end{table}  
                     
As mentioned in the introduction, there are theoretical models which favour
$\omega$-production to proceed via $N^*$-resonances (${\rm pp\to p}N^*$,
$N^*\to \rm p\omega$). The finding of an isotropic, at most slightly anisotropic proton angular distributions at all energies and 
that of undeniably anisotropic ones for $\omega$-mesons at the two higher excess energies are difficult to 
reconcile with resonant $\omega$-production as the major reaction mechanism. This will be shown in the following discussion.

In the present experiment the mass region of 1.721
GeV/c$^2$ (threshold) to 1.893 GeV/c$^2$ (highest beam momentum) was
covered. In this region nucleon resonances are known which have
significant $N^*\to \rm p\omega$ decay branches, namely N(1710)$\rm P_{11}$,
N(1720)$\rm P_{13}$, and N(1900)$\rm P_{13}$ \cite{pdg}.
 Below the threshold N(1535)$\rm S_{11}$,
N(1650)$\rm S_{11}$, N(1675)$\rm D_{15}$, N(1680)$\rm F_{15}$, and N(1700)$\rm
D_{13}$ are listed in  \cite{pdg}.  Most of them are classified as four star resonances, 
while N(1700) and N(1710) have three and N(1900) two stars. Above
the accessible mass range the two star resonances N(1990)$\rm F_{17}$,
N(2000)$\rm F_{15}$, N(2080)$\rm D_{13}$  are located. All these $N^*$
resonances have large widths of typically 100 - 200 MeV and, hence, their
Breit-Wigner tail may well extend into the mass interval covered by this
experiment. The inclusion of subthreshold resonances has proven important in
the calculation of coupling constants of baryon resonances to the
nucleon-omega channel in particular the subthreshold resonances $\rm S_{11}$
and $\rm D_{13}$; above the $\omega$-threshold N(2080)$\rm D_{13}$ was found
to be of particular importance \cite{postNmosel}. It should also be stressed
that the N(1535)$\rm S_{11}$ resonance was specifically claimed to play a
crucial role in off-shell $\omega$-production \cite{Fuchs03}. Hence, in order
to be as unbiased as possible, we include all of these resonances in the
following consideration on resonant $\omega$-production.

In the case of the ${\rm pp \to p}N^*$, $N^*\to\rm p\omega$ reaction the CMS angular distribution of the proton 
associated with the $N^*$-resonance 
depends on spin and parity of the $N^*$-resonance and can be calculated using the formalism 
developed for nuclear reactions by Blatt and Biedenharn \cite{BB}. They showed that the transition from an initial 
state with $J = L_i + S_i$ to a final state with $J = L_f + S_f$
(here the final state is the p$N^*$ system) is characterized by an angular distribution
which can be described by an expansion in even Legendre polynomials $P_L$ in analogy to eq. (\ref{eq:legendrefit}). 
The expansion coefficients are uniquely determined for a given set of quantum numbers $L_i, S_i, J, L_f, S_f$ 
from Wigner and Clebsch-Gordan coefficients.
The maximum of the summation index is $L_{\rm max}$ which, in the case of
reactions with integer values of $J$, is given by $L_{\rm max}$= min$(L_i, J, L_f)$. Examples of angular distributions
for protons from the ${\rm pp} \to {\rm p}N^*(1720)$ reaction for various sets of quantum numbers ($L_i, S_i, J, L_f, S_f$)
can be found in \cite{Faessler04}. 

We used this formalism \cite{BB} to find those ${\rm pp
  \to p}N^*$-reactions which show isotropic proton angular distributions. They are obtained only for transitions from the
initial states $\rm ^1S_0$ or $\rm ^3P_0$ to any $N^*$-resonance and from $\rm ^3P_1$ to
$\rm S_{11}$. This is the consequence of either $L_i, J$ or $L_f$ being
zero, thus $L_{\rm max}=0$. Then, whatever resonance is involved, the angular distribution of the decay products is isotropic.
Thus we have pinned down three initial states which can lead via resonant $\omega$-production
to both isotropic proton and $\omega$ angular distributions.

All other initial states result in angular distributions of various degree of anisotropy in the production step. 
In particular the transitions from
initial states $\rm ^3P_1$ and $\rm ^3P_2$ (amplitudes $f_1$ and $f_7$ of table~\ref{tab:pw}) 
to any $N^*$-resonance (except $\rm ^3P_1$ to $\rm S_{11}$ which is isotropic) give angular distributions of the type $a_0 + a_2\rm P_2$.
If the $N^*$-resonance now decays isotropically, then
the anisotropy from the production process is diluted. As $N^*$-resonances such as $\rm P_{11}$ and $\rm P_{13}$
decay with a relative angular momentum of 1$\hbar$, the anisotropy from the production process may be enhanced. 
This applies to the resonances N(1710)$\rm P_{11}$, N(1720)$\rm P_{13}$, and N(1900)$\rm P_{13}$ 
in the mass region accessible in this experiment. We conclude that resonant $\omega$-production can produce isotropic as well as 
anisotropic angular distributions for protons and $\omega$-mesons, however the degree of anisotropy at $\varepsilon$ = 92MeV
indicates that this contribution is of minor weight. It is also conceivable that resonant production provides a minor contribution at
the higher energies.

The initial states not mentioned explicitly so far yield proton angular distributions from the production step that are of the type 
$a_0 + a_2\rm P_2 + a_4\rm P_4$ and may even include an $a_6\rm P_6$ term. 
This implies also an anisotropic sequential $N^*$-decay, and hence, anisotropic $\omega$ angular distributions
are expected. We are not aware of any mechanism
that would yield isotropic proton angular distributions when superimposing those of anisotropic 
production and anisotropic sequential decay. Hence we conclude that the isotropic proton angular distributions and 
the strongly anisotropic $\omega$ angular distributions
observed at the higher excess energies are incompatible with 
resonant $\omega$-production dominating the reaction mechanism.

We summarise the conclusions drawn from the interpretations of the angular
distributions of $\omega$-mesons and protons in the overall CMS: the former are produced with
exit channel angular momenta of 0, 1, and 2 $\hbar$ depending on excess energy  
while protons are found to be emitted almost isotropically at all excess
energies. This implies that protons are produced most probably in the $\rm ^1S_0$ 
(amplitudes $\rm f_1, f_2, f_3, f_7, f_8, f_9$ of table~\ref{tab:pw}), 
and less probably in the $\rm ^3P$ final state (amplitudes $f_4$, $f_5$, $f_6$). The contribution
of the other amplitudes is assumed to be insignificant. 
The definite anisotropy of the $\omega$ angular distribution and the almost isotropic proton angular distribution
observed at $\varepsilon$ = 128 and 173 MeV rules out resonant $\omega$-production to be the major reaction mechanism
at these excess energies. It cannot be excluded that it contributes to the angular distributions
of $\omega$-mesons and protons to a minor extent at all excess energies as resonant $\omega$-production 
via the $N^*$-resonances $\rm S_{11}$, $\rm D_{13}$ yields isotropic and via
$\rm P_{11}$ or $\rm P_{13}$ anisotropic angular distributions. In these cases, only particular initial states contribute.

\subsection{Orientation of the $\omega$-spin}

Already in 1962 it has been pointed out by Gottfried and Jackson \cite{GJ} that the orientation of the $\omega$-spin gives
valuable information on the reaction mechanism of $\omega$-production. The orientation of the $\omega$-spin is correlated 
with that of the decay plane of the $\omega$ meson. 
The latter can be determined from the pion momenta in the CMS of the decaying $\omega$-meson.
In our experiment the two protons and the charged pions are detected, however the momentum resolution for these fast 
pions  ($\beta\approx 1$) is not sufficient to make direct use of the measured 
momenta such as to calculate the missing 4-momentum 
of the $\pi^{0}$. Nevertheless, its 4-momentum can be determined by exploiting the time of flight of the two observed pions, 
$TOF_{\rm obs,\pi}$, and the $\omega$-momentum. The latter is known as 
$p_{\omega} = p({\rm pp}_{\rm initial})- p({\rm pp}_{\rm final})$ and the
4-momenta of all three  pions must add up to
$p_{\omega}=p_{\pi^-}+p_{\pi^+}+p_{\pi^0}$.  
By varying the 4-momenta of the charged pions and calculating the
corresponding time of flight, $TOF_{\rm calc,\pi}$,
one can find that momentum partition which yields the minimum deviation 
$(\Delta TOF)^2 = (TOF_{\rm calc,\pi_1}-TOF_{\rm obs,\pi_1})^2 +(TOF_{\rm
  calc,\pi_2}-TOF_{\rm obs,\pi_2})^2$.
This method allows to deduce all pion momenta and thus the orientation of the $\omega$ decay plane which is determined 
with an accuracy of $\sigma=10^\circ$.
We like to point out, however, that this analysis cannot be used
to improve on the $\omega$-signal as the background below the signal cannot be reduced.

It was pointed out by Gottfried and Jackson \cite{GJ} and later on by Titov {\it et
al.}\ \cite {Titov2002}, \cite{Titov1999} that it is essential to specify the
reference frame within which an observable is measured. These authors showed that
for the reaction  $\rm pp\to pp\phi, \phi\to \rm K^+ K^-$ the decay angle
$\theta$, defined as the polar angle of the direction of flight of one of the
decay particles in the $\phi$-meson's rest frame with respect to the beam direction, shows an angular
distribution (normalised to 1) given by

\begin{equation}
  {W \rm (cos\theta)}={3\over4}(1-\rho_{00} + (3\rho_{00} - 1)\rm cos^2\theta)
  \label{eq:orientation}
\end{equation}

If $\phi$-production is considered just above threshold the spin of the $\phi$-meson
is aligned and the spin density
matrix element $\rho_{00}=0$; then eq.(\ref{eq:orientation}) reduces to 

\begin{equation}
  {{W \rm (cos\theta)}={3\over4}\rm sin^2\theta}
  \label{eq:rhozero}
\end{equation} 
This relation was used by e.g. Balestra {\it et al.}\ \cite{Balestra2001}, Rekalo {\it et al.}\ \cite{Rekalo} 
and Hartmann {\it et al.}\  \cite{Hartmann} in their investigations of the spin alignment of the $\phi$-meson.  

The appropriate reference axis for the 3-body decay of the $\omega$-meson is the normal to the decay plane. 
The angle of the normal with respect to the 
beam axis becomes $\gamma = \pi/2-\theta$ and replaces the angle $\theta$ in eq.(\ref{eq:orientation})
when translating the 2-body $\phi$-decay into the 3-body $\omega$-decay. Eq.(\ref{eq:orientation}) then reads

\begin{eqnarray}
\nonumber {W(\rm cos\gamma)}&=&\rm N(1-\rho_{00} + (3\rho_{00} - 1)\rm sin^2\gamma)\\
                             &=&\rm N(2\rho_{00} - (3\rho_{00} - 1)\rm cos^2\gamma)
  \label{eq:gammaorientation}
\end{eqnarray}
with the proper normalisation constant N = 1/($2\rho_{00}$+2/3).

We will exploit the eq.(\ref{eq:gammaorientation}) in order to investigate a possible spin alignment in the $\omega$-production. 
The term alignment arises from the following consideration: 
the orbital angular momentum of the initial state of the $\rm pp\to pp\omega$ reaction is
always oriented perpendicularly to the beam and the projection quantum number
of the total angular momentum $M_{J_i}$ of the pp-system can only assume the
values of 0 or $\pm1$ as the Clebsch-Gordan coefficients show. The restriction
$M_{J_i} = 0, \pm1$ carries over to the final state, and the notion of an
aligned total angular momentum applies if $M_{J_f} = \pm1$ only.  

Directly at threshold, the transition $\rm^3P_1\to {^1S_0}s$ (amplitude $f_1$
of table~\ref{tab:pw}) is the only possible one. In this case, $J_i = J_f =
s_{\omega}=1$ and $M_{J_i}=m_{s_{\omega}}=\pm1$, i.e. the spin of the
$\omega$-meson is aligned (which goes along with $\rho_{00} = 0$) and the
angular distribution becomes according to eq. (\ref{eq:gammaorientation})

\begin{equation}
  {{W \rm (cos\gamma)}={3\over2}\rm cos^2\gamma}
  \label{eq:omega-rhozer}
\end{equation} 

This finding, solely based on conservation laws which are reflected by the appropriate Clebsch-Gordan
coefficients, is true independent of the reaction model under consideration. 

It should be pointed out, that $\rho_{00} = 0$ holds only at threshold but
assumes values of about 0.2 already slightly above. This is a consequence of the spin-orbit interaction
\cite{Titov2003}. Naturally, this effect increases with the vector meson momentum $\vec q$.  

Above threshold, we have to consider firstly, Sp type final states (amplitudes $f_2$
and $f_3$ of table~\ref{tab:pw}). For the amplitude $f_2$ the initial state
$\rm^1S_0$ does not exhibit any orientation, hence the $\omega$-spin has no
particular orientation and $W(\rm cos\gamma) = {1/2}$ is isotropic. This
reflects the fact that all spin projection states are equally populated and, hence, the spin density matrix element
$\rho_{00}=\rho_{11}=\rho_{-1-1}$=1/3. For the amplitude $f_3$ ($\rm^1D_2
\to {^1S_0}p$) one finds $M_{J_f}=0$, i.e. the total angular momentum of the
$\omega$-meson is perpendicular to the beam axis. The $\omega$-spin projection
quantum number, however, can assume the values $m_{s_{\omega}}=0,\pm1$ resulting in a
distribution of the type $a \cdot {\rm sin^2\gamma} + b \cdot {\rm
cos^2\gamma}$ which tends to reduce the anisotropy due to amplitude $\rm f_1$.
Secondly, we consider Sd type final states. Distributions of the same type as for amplitude $f_3$ 
are expected. Thirdly, if we take into account Ps type final states we find isotropy for amplitude $\rm f_4$, 
type $a \cdot {\rm sin^2\gamma} + b \cdot {\rm cos^2\gamma}$ distributions for $\rm f_5$ and alignment for $\rm f_6$.

\begin{figure*}
\resizebox{1.\textwidth}{!}{ \includegraphics{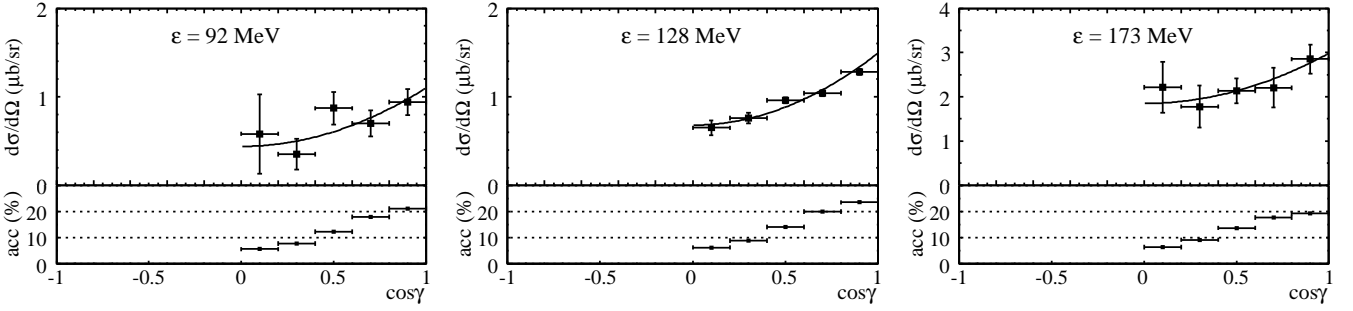} }
\caption{Distributions of the angle between the direction of the $\omega$-spin
        and the beam axis at the three excess energies (upper part in each frame; lower part: acceptance).
      See text for details.}
\label{fig:dpresz}  
\end{figure*}

The measured angular distribution of the $\omega$-spin direction is shown in
fig.~\ref{fig:dpresz}. We describe these angular distributions by employing eq.(\ref{eq:gammaorientation})

\begin{equation}
  {{\rm d}\sigma\over{\rm dcos}\gamma}={\sigma_{tot}\over 2\rho_{00}+2/3} (2\rho_{00} - (3\rho_{00} - 1)\rm cos^2\gamma).
  \label{eq:gammaorientation2}
\end{equation}

Fig.~\ref{fig:dpresz} depicts the results of the fitting as a solid line. 
The values determined for $\sigma_{tot}$ and ($\rho_{00}$) are listed in
table~\ref{tab:dpresz}. 

\begin{table}
  \caption{Parameters of the 
      function eq.(\ref{eq:gammaorientation2}) used to describe the distribution of the angle $\gamma$ between the
      normal on the $\omega$-decay plane and the beam-axis. }
    \label{tab:dpresz}     
    \begin{tabular}{@{}rrr}     
      \hline\noalign{\smallskip}                           
      $\varepsilon\;(\mathrm{MeV})$ & $\rm \sigma_{tot}\;(\mu b)$ & $\rho_{00} $   \\
      \noalign{\smallskip}\hline\noalign{\smallskip}
     $92$ & $ 8.3 \pm 1.2$ & $0.17 \pm 0.07$ \\
    $128$ & $12.0 \pm 0.6$ & $0.19 \pm 0.03$ \\
    $173$ & $28.0 \pm 2.3$ & $0.24 \pm 0.05$ \\
      \noalign{\smallskip}\hline                       
    \end{tabular}                 
\end{table}  
             
The increasing value of $\rho_{00}$ indicates the transition from an aligned $\omega$-spin ($\rho_{00}=0$) to an
arbitrarily oriented  $\omega$-spin ($\rho_{00}=1/3$). This is due to two effects, firstly the increasing importance of
reaction amplitudes that show no alignment, namely those of the initial state $\rm^1S_0$ which provide an isotropic contribution to 
the angular distribution as well as those which feature a $\rm sin^2\gamma$-dependence, and secondly the 
the above mentioned increase of $\rho_{00}$ due to spin orbit interaction being important above threshold
\cite{Titov2003}. The energy dependence of the parameter $\rho_{00}$ sets a benchmark for theoretical models 
of $\omega$-production. 

\subsection{Invariant mass spectra}

Invariant mass spectra of the two-body subsystems can be used to search for deviations from phase space 
which would be indicative of a resonance. In order to determine an invariant mass spectrum
the $\omega$-signal needs to be extracted from the missing mass spectra generated for various bins in the invariant mass. 
It turned out that this was possible only
if a Monte Carlo simulated pion-cocktail was used for the determination of the background. The statistical basis of our data does not
allow to generate a Dalitz plot.

\begin{figure*}
\resizebox{1.\textwidth}{!}{ \includegraphics{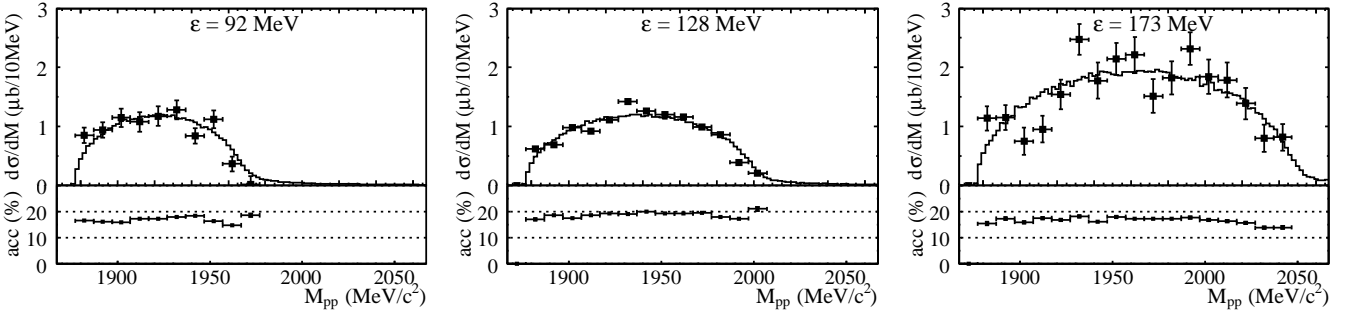} }
\caption{Invariant pp-mass distribution (upper part of each frame, acceptance: lower part). The histograms 
      represent the phase space distributions.}
\label{fig:invmpp}  
\end{figure*}

\begin{figure*}
\resizebox{1.\textwidth}{!}{ \includegraphics{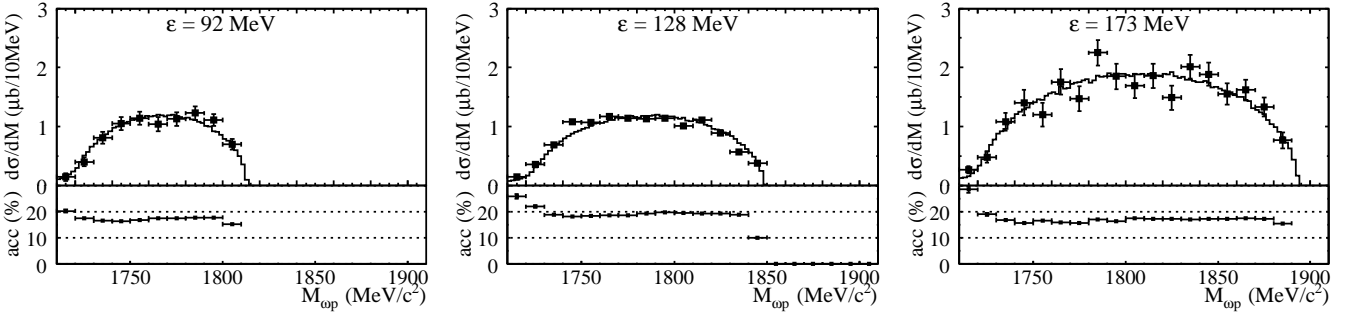} }
\caption{Invariant $\omega$p-mass distribution (upper part of each frame, acceptance: lower part). 
      The histograms represent the phase space distributions.}
\label{fig:invmwp}  
\end{figure*}

The results obtained are shown in figs.~\ref{fig:invmpp} and \ref{fig:invmwp}
(upper part of each frame, the lower part shows the acceptance) together with
the simulated distribution based on an equally populated phase space for the $\omega$-production. 
The fuzzy fringes of the simulated distributions (upper end of $\rm
M_{\rm pp}$ and lower end of $\rm M_{\rm \omega p}$) reflect the finite width of the $\omega$-meson; the
experimental data are reproduced very well. $N^*$-resonances with widths in the
order of 100 to 200 MeV do not cause the invariant mass spectrum to deviate
perceptibly from phase space, as simulations have shown. Arguing the other
 way around we conclude that any resonance with a width
below our missing mass resolution of 10 MeV would show up if their cross
section would be above $\approx$ 0.5$\mu$b. 

It is obvious from these results that indications of resonant
$\omega$-production are missing. This finding is compatible with that from the
investigation of the proton angular distributions. An observable closely related
to the invariant mass will be discussed in the following subsection. 
\subsection{Helicity and Jackson angle} 

\subsubsection{Notation}

In a $2\to 3$ reaction $(a + b \to 1 + 2 + 3)$ with conservation of 4-momentum
given by 
\begin{equation}
  p_a  + p_b = p_1 + p_2 + p_3
\label{eq:pconservation}
\end{equation}
the initial state in the overall CMS is defined by a direction and the total
energy $\sqrt{s}$ which both are known in standard collision experiments. The
description of the final state requires $3 \times 3 = 9$ variables (besides
the masses in the exit channel) which, due to conservation of momentum and
energy,  reduce to five variables. Without polarisation in the entrance channel the
azimuthal orientation of the final state is arbitrary, reducing the number of
necessary variables to four.

The four variables needed to uniquely describe the reaction can be
represented by two invariants and two angles. For the latter one may choose e.g. the CMS angles of any two final state particles;
for the  invariants one often chooses the invariant masses of 2-body subsystems
\begin{eqnarray}
s_{13} = (p_1 + p_3)^2 = (p_a + p_b - p_2)^2 \\ 
\label{eq:invs13} 
s_{23} = (p_2 + p_3)^2 = (p_a + p_b - p_1)^2 
\label{eq:invs23} 
\end{eqnarray}
where $s_{12}$ may be chosen as an alternative to $s_{13}$ or $s_{23}$. With this choice, only quantities of the exit channel are 
taken into account. This shortcoming can be avoided by choosing a suitable angle in an appropriate Lorentz reference system \cite{GJ}.
The latter is obtained by boosting the final state in such a way
that the momenta of particle 2 and 3 add up to zero; one then obtains from eq. (\ref{eq:pconservation}) 
\begin{equation}
  0 = \vec p_2 + \vec p_3 = \vec p_a + (\vec p_b - \vec p_1) = \vec p_b + (\vec p_a - \vec p_1).
  \label{eq:GJframe}
\end{equation}

Now, an angle connecting exit and entrance channel is the (polar) Jackson angle (notation of Byckling and Kajantie \cite{Byckling}) 
\begin{equation}
  \cos\theta^{\mathrm{R}23}_{b3} = {\vec p_b\cdot \vec p_3\over |\vec p_b|\cdot| \vec p_3|}\Bigr|_{\vec p_2=-\vec p_3}.
  \label{eq:jacksonangle}
\end{equation}
The superscript specifies the Lorentz reference system R23 chosen through $\vec p_2=- \vec p_3$; the first subscript 
indicates which particle 
defines via its momentum direction the polar reference axis (a or b for
the Jackson frame). The second subscript specifies which particle is used when calculating the
Jackson angle. Since $\vec p_2=- \vec p_3$, the choice of this
particle is arbitrary, and $\theta^{\mathrm{R}23}_{b3} = \pi -
\theta^{\mathrm{R}23}_{b2}$. 

In the same Lorentz reference system one can also calculate the (polar) helicity angle
\begin{equation} 
  \cos\theta^{\mathrm{R}23}_{13} = {\vec p_1\cdot \vec p_3\over |\vec p_1|\cdot|\vec p_3|}\Bigr|_{\vec p_2=- \vec p_3}. 
  \label{eq:helicity angle}
\end{equation}
which, in contrast to the Jackson angle depends on exit channel properties only.
Again the Lorentz reference system "R23" chosen through $\vec p_2=- \vec p_3$ is specified by the superscript, while the first  
subscript indicates which particle 
defines via its momentum direction the polar axis (1 for the helicity frame). 
The second subscript specifies which particle is used when calculating the
helicity angle. Since $\vec p_2=- \vec p_3$, the choice of this
particle is arbitrary, and $\theta^{\mathrm{R}23}_{13} = \pi -
\theta^{\mathrm{R}23}_{12}$. Of course, all formulae are also valid for any
other Lorentz reference systems (superscript R12 or R13 in eq.(\ref{eq:jacksonangle}) or eq.(\ref{eq:helicity angle})
with the corresponding first and second subscript. 

\subsubsection{Helicity angle}

\begin{figure*}
\resizebox{1.\textwidth}{!}{ \includegraphics{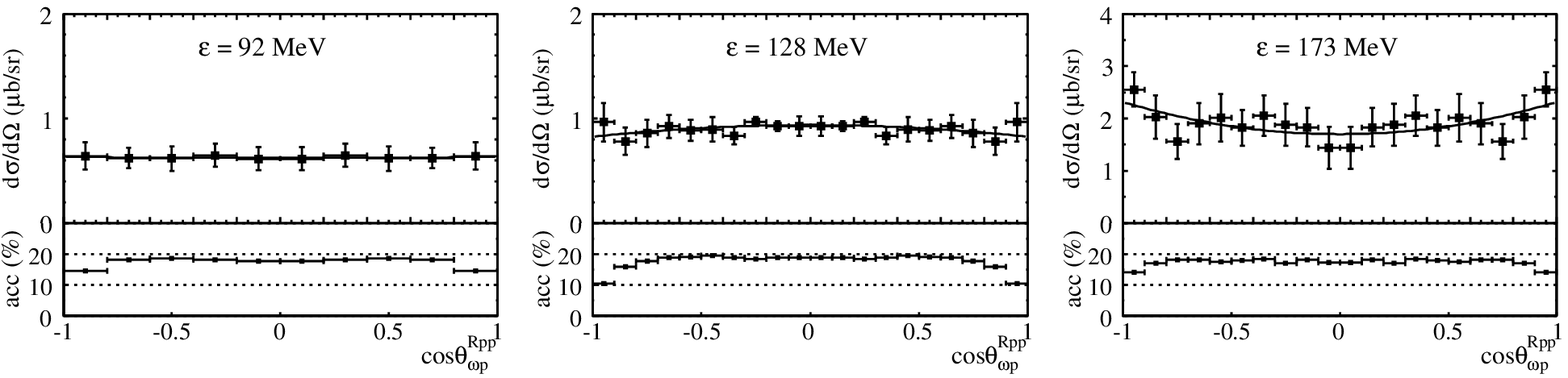} }
\caption{Helicity angle distribution of the $\omega$-meson in the pp-helicity frame
      (upper part of each frame, acceptance: lower part).}
\label{fig:finhelicitypp}  
\resizebox{1.\textwidth}{!}{ \includegraphics{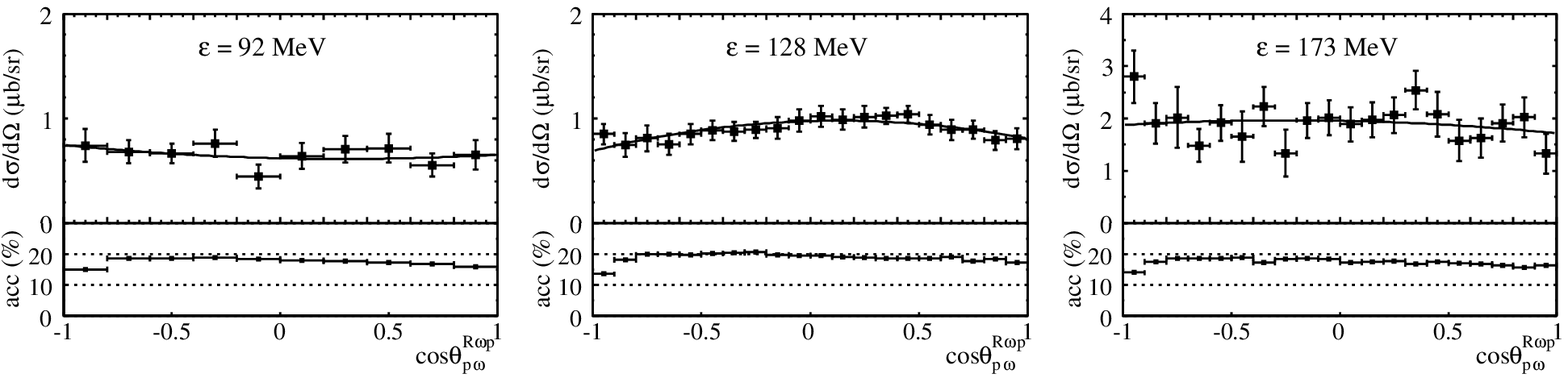} }
\caption{Helicity angle distribution of the "other" proton in the
      $\omega$p-helicity frame (upper part of each frame, acceptance:
      lower part).}
\label{fig:finhelicitywp}  
\resizebox{1.\textwidth}{!}{ \includegraphics{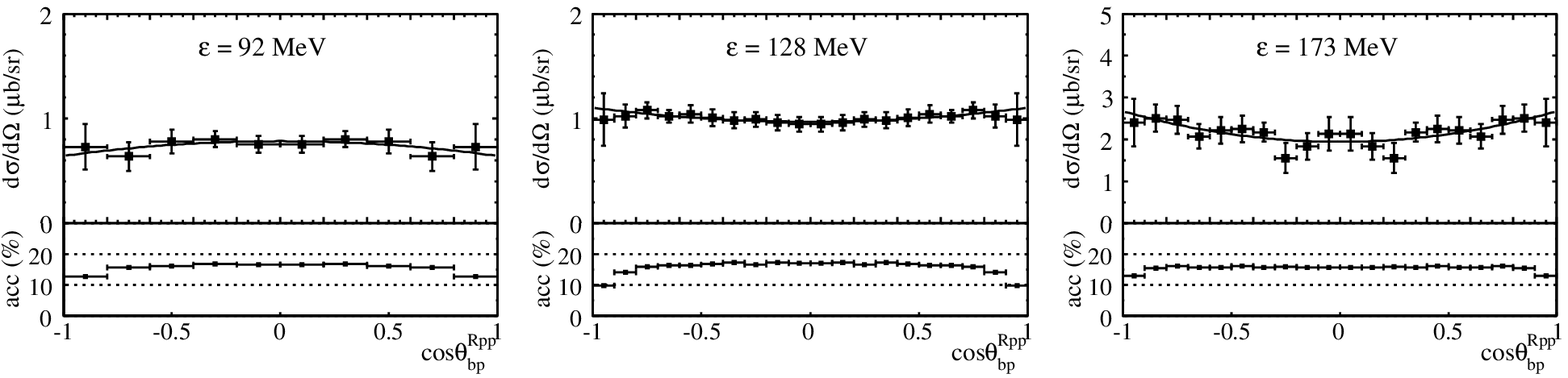} }
\caption{Distribution of the angle between the beam direction and that of
      a proton in the pp-Jackson frame (Jackson angle) (upper part of each frame,
    acceptance: lower part).}
\label{fig:finjacksonpp}  
\resizebox{1.\textwidth}{!}{ \includegraphics{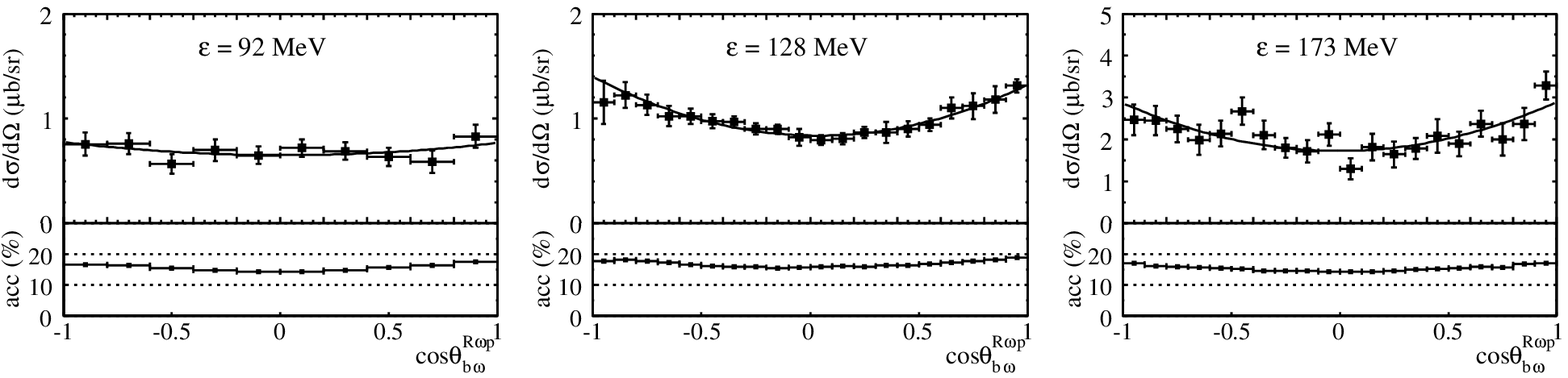} }
\caption{Distribution of the angle between beam and the $\omega$-meson in
      the $\omega$p-Jackson frame (Jackson angle) (upper part of each frame,
    acceptance: lower part).}
\label{fig:finjacksonwp}  
\end{figure*}

The helicity angle relates particles 2 and 3 to particle 1, and hence carries
information solely about the exit channel. This information is also contained in the
Dalitz-plot: A resonance in the 2-3-system is observed as an enhancement of
events along the $s_{23}$-axis at a certain mass and the distribution of
$\cos\theta^{R23}_{13}$ is simultaneously seen 
along the $s_{13}$-axis. From this helicity angle distribution
the spin of the resonance can be inferred in particular cases. It should be
noted that this angular distribution is not necessarily symmetric
with respect to $\cos\theta^{R23}_{13}=0$ as it does not relate to the entrance channel. 
   
Two helicity angles can be considered for the reaction $\rm pp \to pp\omega$:
In the first case the two protons are used to define the Lorentz reference system
and the polar helicity angle of the $\omega$-meson is determined with respect to one
of the protons. Since the two protons are indistinguishable we
must take into account two distributions, namely $\theta^{\rm Rpp}_{\rm
  \omega p_1}$ and $\theta^{\rm Rpp}_{\rm \omega p_2}$. This results in a  
symmetrised helicity angle distribution.

The result is shown in fig.~\ref{fig:finhelicitypp} where all possible
values of $s_{12}$ are accounted for. This corresponds to a summation of
Dalitz-plot entries with constant values of $\cos\theta^{\rm Rpp}_{\rm
  \omega p}$. The solid lines represent the results of least square fitting using eq.(\ref{eq:legendrefitsigma}).
 The coefficients are listed in the first section of table~\ref{tab:coeffhJ}.

At $\varepsilon=$ 92 MeV one observes isotropy as well as at $\varepsilon=$ 128 MeV. At
$\varepsilon$ = 173 MeV the increase of cross section at $\rm cos\theta^{\rm
  Rpp}_{\rm \omega p}=\pm1$ seems to indicate that the $\omega$-meson has some tendency
to move along with a proton. This may reflect the angular momentum of the
$\omega$-meson with respect to the $\rm ^1S_0$ di-proton and not necessarily
a contribution of a resonance. A theoretical model is asked for to explain
this observation.

\begin{table*}
  \caption{Coefficients of Legendre polynomials
    determined by fitting eq.(\ref{eq:legendrefitsigma}) to the distributions of the
    different observables.}
  \label{tab:coeffhJ}
  \begin{tabular}{@{}rrrrrr}      
      \hline\noalign{\smallskip}                            
      observable & $\varepsilon\;(\mathrm{MeV})$ & $\sigma_{\rm tot}\;(\mathrm{\mu b})$  & $a'_1$ & $a'_2$ \\
      \noalign{\smallskip}\hline\noalign{\smallskip}
      $\cos\theta^{\rm Rpp}_{\rm\omega p}$ &  $92$ & $ 8.8 \pm 0.5$ & -- & $0.01 \pm 0.14$   \\ 
      (fig. \ref{fig:finhelicitypp})       & $128$ & $12.7 \pm 0.3$ & -- & $-0.08 \pm 0.07$ \\
                                           & $173$ & $26.8 \pm 1.2$ & -- & $0.22 \pm 0.10$   \\
      \noalign{\smallskip}\hline\noalign{\smallskip}
      $\cos\theta^{\rm R\omega p}_{\rm p\omega}$ &  $92$ & $ 9.1 \pm 0.5$ & $-0.07 \pm 0.11$ & $0.08 \pm 0.15$ \\ 
      (fig. \ref{fig:finhelicitywp})             & $128$ & $12.7 \pm 0.3$ & $0.07 \pm 0.04$ & $-0.17 \pm 0.06$ \\
                                                 & $173$ & $26.8 \pm 1.2$ & $-0.04 \pm 0.08$ & $-0.06 \pm 0.11$ \\
      \noalign{\smallskip}\hline\noalign{\smallskip}                             
      $\cos\theta^{\rm Rpp}_{\rm bp}$      &  $92$ & $ 9.2 \pm 0.5$ & -- & $-0.13 \pm 0.15$ \\ 
      (fig. \ref{fig:finjacksonpp})        & $128$ & $12.7 \pm 0.2$ & -- & $0.09 \pm 0.05$  \\
                                           & $173$ & $27.5 \pm 0.9$ & -- & $0.22 \pm 0.09$  \\
      \noalign{\smallskip}\hline\noalign{\smallskip}
      $\cos\theta^{\rm R\omega p}_{\rm b\omega}$ &  $92$ & $ 8.7 \pm 0.4$ & $-0.01 \pm 0.08$ & $0.12 \pm 0.11$ \\ 
      (fig. \ref{fig:finjacksonwp})              & $128$ & $12.7 \pm 0.2$ & $-0.04 \pm 0.04$ & $0.35 \pm 0.04$ \\
                                                 & $173$ & $26.5 \pm 0.9$ & $0.01 \pm 0.06$ & $0.37 \pm 0.08$ \\
     \noalign{\smallskip}\hline
    \end{tabular}                 
\end{table*}                       
 
In the second case one proton (for instance particle 2) and the $\omega$-meson
(particle 3) are used to define the Lorentz reference system, and the polar
helicity angle of the other proton (particle 1) is determined with respect to
the direction of the $\omega$-meson. Since the two protons are
indistinguishable we must average the distributions of $\theta^{\rm R\omega
  p_1}_{\rm p_2\omega}$ and $\theta^{\rm R\omega p_2}_{\rm p_1\omega}$ for a
given event. A possible anisotropy in the helicity angle is not destroyed by
this procedure. 

Fig.~\ref{fig:finhelicitywp} shows this helicity angle distribution where
all possible values of $s_{23}$ were taken into account. This corresponds to a
summation of Dalitz-plot entries with constant values of $\cos\theta^{\rm
  R\omega p}_{\rm p\omega}$. The solid lines represent the results of least square fitting
using eq.(\ref{eq:legendrefitsigma}) but allowing  $\rm P_1$ in addition with a weight of $a'_1=a_1$/$a_0$.
The coefficients are listed in the second section of table~\ref{tab:coeffhJ}.   

The helicity angle distributions at $\varepsilon=$ 92 and 173 MeV are isotropic within uncertainties.
At $\varepsilon =$ 128 MeV significant anisotropy and asymmetry values are observed pointing at an inhomogeneous population of
the Dalitz plot which is probably caused by the angular momentum in the exit channel.
This finding, however, is not in contradiction with the
helicity angle distribution of fig.~(\ref{fig:finhelicitypp}) as a different
projection of the Dalitz plot is presented. Again, without a theoretical model
one cannot explain the origin of this effect.

\subsubsection{Jackson angle}

The Jackson angle relates the direction of the beam momentum with
the orientation of the axis given by $\vec p_2=- \vec p_3$.
Eq.(\ref{eq:GJframe}) can be interpreted, as suggested by Gottfried and Jackson
\cite{GJ}, as a $2 \to 2$ reaction, namely $2 + 3 \to b + x$  where particle x
with $p_x = p_a - p_1$ can be identified with the exchange meson in diagram
(a) or (c) in fig.~\ref{fig:diagram}. Thus it is clear that any
structure in the Jackson angle distribution gives direct information on
the angular momentum of the system R23 which could, but not necessarily has to,
be a resonance \cite{Sibirtsev06}. 

Two Jackson angles can be considered for the reaction $\rm pp \to pp\omega$:
In the first case the two protons (particle 1 and 2) are used to define the
Lorentz reference system and the angle between the direction of one of the two
protons with respect to the beam direction is called the polar Jackson
angle. Since the protons are indistinguishable we must take into account the
angles $\theta^{\rm Rpp}_{\rm bp_1}$ and $\theta^{\rm Rpp}_{\rm
  bp_2}$, a procedure which leads to a distribution of a symmetrised polar
Jackson angle $\theta^{\rm Rpp}_{\rm bp}$.
Since beam and target proton can not been distinguished, both Jackson angles
with respect to beam and target are taken into account for each
event (maintaining for simplicity the sudscript $b$ in $\theta^{\rm Rpp}_{\rm
  bp}$ )

In fig.~\ref{fig:finjacksonpp} we show this symmetrised Jackson angle
distribution. The solid lines represent the results of least square fitting using eq.(\ref{eq:legendrefitsigma}). 
The coefficients are listed in the third section of table~\ref{tab:coeffhJ}.

The fits suggest a slight anisotropy which implies that there is at most a very weak 
correlation between the beam direction and the final proton
pair. This is compatible with the proton angular distributions presented in fig.~\ref{fig:cosp} which showed an
insignificant contribution of Ps type final states. 

In the second case one proton (for instance particle 1) and the $\omega$-meson
(particle 3) are used to define the Lorentz reference system, and the polar
Jackson angle is measured as the direction of the $\omega$-meson with respect to the
beam direction. Since the two protons in the exit channel are
indistinguishable we must average the angles $\theta^{\rm R\omega p_1}_{\rm
  b\omega}$ and $\theta^{\rm R\omega p_2}_{\rm b\omega}$ in order to yield
$\theta^{\rm R\omega p}_{\rm b\omega}$. This averaging causes a dilution of a
signal from a potential resonance since this is either found in the $\rm
\omega p_1$ or $\rm \omega p_2$ system, while the non-resonant one furnishes
an uncorrelated, hence isotropic background. 

Again, the indistinguishability of beam and target are taken into account.
However, in the Lorentz reference system beam and target momenta are not aligned,
hence this procedure is not a symmetrisation,
but the distribution turns out to be symmetric.

In fig.~\ref{fig:finjacksonwp} we show this Jackson angle distribution. The
solid lines represent the results of least square fitting using eq.(\ref{eq:legendrefitsigma}). 
The coefficients are listed in the fourth section of table~\ref{tab:coeffhJ}.

\begin{figure*}
\resizebox{1.\textwidth}{!}{ \includegraphics{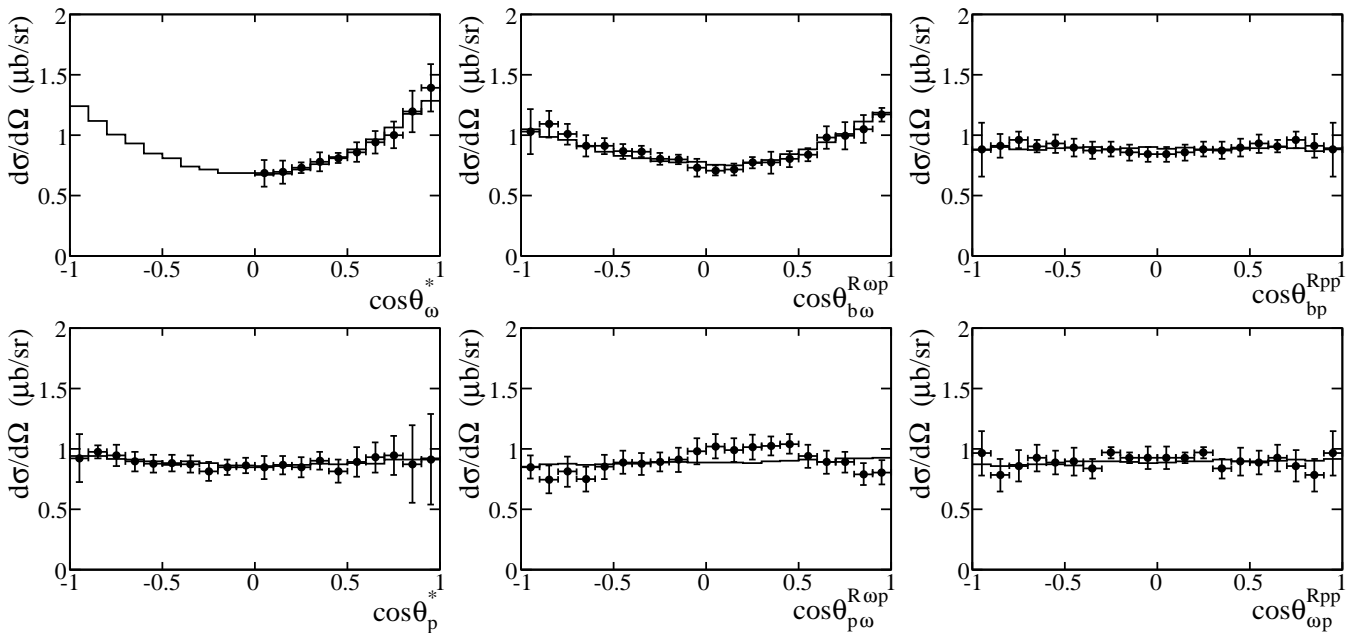} }
\caption{Comparison of the experimental results with a Monte Carlo simulation using
      the measured distribution of $\rm cos\theta_{\omega}$ as a weight function.}
\label{fig:filtercosw}  
\end{figure*}

The Jackson angle distribution at $\varepsilon$ = 92 MeV is almost isotropic but
pronounced anisotropies are observed at the other excess energies beyond
doubt.
This shows that there is a non-zero relative angular momentum in the $\omega$p
system. However, a conclusion that this is a direct signal of a
$N^*$-resonance decaying into p$\omega$ should be drawn with caution. The
deduced values of $L=0,1,2$, as discussed in the context of the $\omega$
angular distributions (fig.~\ref{fig:cosw}), are reflected by the Jackson
angular distribution.
In order to corroborate this argument we present in fig.~\ref{fig:filtercosw}
(together with data already shown) results of Monte Carlo
simulations at $\varepsilon$ = 128 MeV where the
event generator was modified such as to reproduce the $\omega$ angular
distribution of fig.~\ref{fig:cosw}, again shown in the upper left frame of fig.~\ref{fig:filtercosw}.
This weight function on the $\omega$ angular distribution also
modifies the proton angular (lower left) and Jackson angle distributions (upper middle and right frame)
in such a way that
they perfectly match the experimental ones. We obviously observe in the
Jackson angle distribution the deviation of the reaction kinematics from pure phase
space. The helicity angle distributions (lower middle and lower right frame), however, cannot be
reproduced.

We tentatively conclude that we observe the influence of some particular
reaction dynamics: the transition matrix element connecting the initial with
the final state shows a dependence on $\vec q$ and $l_{\omega}$ but not on $\vec
p$ or $L$, creating anisotropic angular distributions of the $\omega$-mesons,
helicity angle $\theta^{\rm  R\omega p}_{\rm p\omega}$, and Jackson angle
$\theta^{\rm  R\omega p}_{\rm b\omega}$. The final state protons are in the
$\rm ^1S_0$ state, resulting in isotropic angular distributions of $\theta^*_{\rm p}$,
helicity angle $\theta^{\rm  Rpp}_{\rm \omega p}$, and Jackson angle
$\theta^{\rm  Rpp}_{\rm bp}$.  

We would like to note that the results of the corresponding investigation at 
$\varepsilon$ = 173 MeV allow the same conclusion.

\section{Summary}

In this paper we presented a systematic study of the production of $\omega$-mesons in
proton-proton-collisions, carried out in a kinematically complete experiment at three excess energies of
$\varepsilon= 92, 128, 173$ MeV. The large-acceptance COSY-TOF spectrometer
allows the almost unambiguous and simultaneous identification of different reaction channels. 
We described in detail how the yield of $\omega$-events was determined in
the presence of an unavoidable physical background which is due to 
the production and decay of the $\rho$-meson ($\rho \to \pi^+\pi^-$) and non-resonant $\pi^+\pi^-$
as well as $\pi^+\pi^-\pi^0$ production. 
Total cross sections, angular distributions both of
$\omega$-mesons and protons were measured in the overall CMS as well as helicity and Jackson
angle distributions in both the pp and $\omega$p helicity and Jackson
frames, respectively. In addition, the orientation of the $\omega$-spin and invariant mass
spectra were determined. All total cross sections obtained by integrating the various differential distributions agree within
uncertainty, proving the consistency of our data evaluation.
  
  The major results are as follows: We conclude from the angular distribution of the $\omega$-meson
that its production takes place
  dominantly in Ss and Sp final states for the lower energies, where only little room is left for Ps final states,
and, additionally, type Sd final states for the highest energy. The production of $\omega$-mesons via
  $N^*$-resonances was ruled out to be the major reaction mechanism. It is, however, conceivable that resonant
 $\omega$-production via the broad subthreshold
  resonances $\rm S_{11}$ and $\rm D_{13}$ as well as resonances above threshold, $\rm P_{11}$ and $\rm P_{13}$, 
may happen for particular initial states. However it probably plays a minor
role among the various $\omega$ production mechanisms discussed in the literature.
Invariant mass spectra for both the pp and p$\omega$ subsystems are found to
be compatible with phase space distributions underlining the minor importance
of resonant $\omega$-production. A dominant role of $\rm ^3P_1$ and $\rm
^1S_0$ initial partial waves for $\omega$-production was concluded from the
orientation of the $\omega$-spin. Although we observe anisotropic Jackson
angle distributions in the $\omega$p-Jackson frame we argue that this is not
an indication of a resonance but rather a kinematical effect reflecting the 
anisotropy of the $\omega$ angular distribution in the overall CMS.
The helicity angle distribution in the $\omega$p-helicity frame shows an
anisotropy which, in addition to the orientation of the $\omega$-spin, is
probably the most sensitive observable to judge the validity of various
theoretical descriptions of the production process.  

\begin{acknowledgement}
The authors would like to express their gratitude to the COSY staff for the
operation of the accelerator during these experiments. Fruitful discussions
with C. Hanhart, W. Gl\"ockle, B. K\"ampfer, A. A. Sibirtsev, and A. I. Titov are
gratefully acknowledged. This work was supported in part by grants from BMBF(DD117I)
and COSY-FFE (Forschungszentrum J\"ulich). 
\end{acknowledgement}

%
%


\appendix
\section*{Appendix}
\begin{table*}[h]
  \caption{Differential cross sections in $\rm\mu b/sr$ of the angular
    distributions. The central value of a cosine interval of 0.2 or 0.1 is listed.}
  \label{tab:diffang}
  \begin{tabular}{@{}lrllllll}      
    \hline\noalign{\smallskip}                            
    & $\cos\theta$ & $\rm {d\sigma\over d\Omega}(\theta_\omega)$ & $\rm {d\sigma\over d\Omega}(\theta_{p})$ &
    $\rm {d\sigma\over d\Omega}(\theta^{Rpp}_{\omega p})$ & $\rm {d\sigma\over d\Omega}(\cos\theta^{R\omega p}_{p\omega})$ &
    $\rm {d\sigma\over d\Omega}(\cos\theta^{Rpp}_{bp})$ & $\rm {d\sigma\over d\Omega}(\cos\theta^{R\omega p}_{ b\omega})$\\
    \noalign{\smallskip}\hline\noalign{\smallskip}
    $\varepsilon = 92\;\mathrm{MeV}$
    & $-0.9$ &                 & $0.69 \pm 0.23$ & $0.72 \pm 0.15$ & $0.83 \pm 0.18$ & $0.73 \pm 0.22$ & $0.75 \pm 0.11$\\
    & $-0.7$ &                 & $0.84 \pm 0.15$ & $0.69 \pm 0.11$ & $0.76 \pm 0.12$ & $0.64 \pm 0.14$ & $0.76 \pm 0.10$\\
    & $-0.5$ &                 & $0.50 \pm 0.12$ & $0.69 \pm 0.13$ & $0.74 \pm 0.11$ & $0.78 \pm 0.12$ & $0.57 \pm 0.10$\\
    & $-0.3$ &                 & $0.73 \pm 0.15$ & $0.73 \pm 0.13$ & $0.85 \pm 0.15$ & $0.80 \pm 0.08$ & $0.70 \pm 0.10$\\
    & $-0.1$ &                 & $0.70 \pm 0.13$ & $0.69 \pm 0.12$ & $0.50 \pm 0.13$ & $0.75 \pm 0.08$ & $0.65 \pm 0.08$\\
    & $ 0.1$ & $0.72 \pm  0.11$ & $0.77 \pm 0.10$ & $0.69 \pm 0.12$ & $0.72 \pm 0.14$ & $0.75 \pm 0.08$ & $0.72 \pm 0.09$\\
    & $ 0.3$ & $0.56 \pm  0.21$ & $0.42 \pm 0.13$ & $0.73 \pm 0.13$ & $0.79 \pm 0.14$ & $0.80 \pm 0.08$ & $0.69 \pm 0.08$\\
    & $ 0.5$ & $0.54 \pm  0.19$ & $0.47 \pm 0.24$ & $0.69 \pm 0.13$ & $0.80 \pm 0.15$ & $0.78 \pm 0.12$ & $0.63 \pm 0.09$\\
    & $ 0.7$ & $0.86 \pm  0.21$ & $0.79 \pm 0.24$ & $0.69 \pm 0.11$ & $0.62 \pm 0.12$ & $0.64 \pm 0.14$ & $0.59 \pm 0.11$\\
    & $ 0.9$ & $0.88 \pm  0.24$ & $0.89 \pm 0.41$ & $0.72 \pm 0.15$ & $0.73 \pm 0.16$ & $0.73 \pm 0.22$ & $0.83 \pm 0.11$\\
    \noalign{\smallskip}\hline\noalign{\smallskip}
$\varepsilon = 128\;\mathrm{MeV}$ 
 & $-0.95$ &                 & $1.03 \pm 0.22$ & $1.08 \pm  0.20$ & $0.95 \pm  0.11$ & $0.99 \pm  0.25$ & $1.15 \pm  0.21$\\
 & $-0.85$ &                 & $1.09 \pm 0.06$ & $0.88 \pm  0.15$ & $0.84 \pm  0.13$ & $1.02 \pm  0.11$ & $1.22 \pm  0.12$\\
 & $-0.75$ &                 & $1.06 \pm 0.10$ & $0.96 \pm  0.15$ & $0.91 \pm  0.14$ & $1.08 \pm  0.08$ & $1.13 \pm  0.10$\\
 & $-0.65$ &                 & $1.00 \pm 0.09$ & $1.04 \pm  0.12$ & $0.84 \pm  0.11$ & $1.02 \pm  0.06$ & $1.02 \pm  0.10$\\
 & $-0.55$ &                 & $0.99 \pm 0.08$ & $0.99 \pm  0.11$ & $0.95 \pm  0.11$ & $1.04 \pm  0.09$ & $1.02 \pm  0.08$\\
 & $-0.45$ &                 & $0.99 \pm 0.08$ & $1.00 \pm  0.13$ & $1.00 \pm  0.11$ & $1.01 \pm  0.08$ & $0.97 \pm  0.07$\\
 & $-0.35$ &                 & $0.98 \pm 0.08$ & $0.94 \pm  0.09$ & $0.98 \pm  0.10$ & $0.98 \pm  0.08$ & $0.97 \pm  0.05$\\
 & $-0.25$ &                 & $0.91 \pm 0.09$ & $1.09 \pm  0.05$ & $1.00 \pm  0.09$ & $0.99 \pm  0.07$ & $0.90 \pm  0.05$\\
 & $-0.15$ &                 & $0.95 \pm 0.07$ & $1.04 \pm  0.05$ & $1.02 \pm  0.11$ & $0.96 \pm  0.07$ & $0.90 \pm  0.04$\\
 & $-0.05$ &                 & $0.97 \pm 0.07$ & $1.04 \pm  0.10$ & $1.10 \pm  0.12$ & $0.94 \pm  0.07$ & $0.82 \pm  0.08$\\
 & $ 0.05$ & $0.77 \pm 0.12$ & $0.95 \pm 0.11$ & $1.04 \pm  0.10$ & $1.14 \pm  0.11$ & $0.94 \pm  0.07$ & $0.79 \pm  0.05$\\
 & $ 0.15$ & $0.78 \pm 0.11$ & $0.97 \pm 0.09$ & $1.04 \pm  0.05$ & $1.11 \pm  0.11$ & $0.96 \pm  0.07$ & $0.80 \pm  0.05$\\
 & $ 0.25$ & $0.82 \pm 0.05$ & $0.95 \pm 0.09$ & $1.09 \pm  0.05$ & $1.14 \pm  0.12$ & $0.99 \pm  0.07$ & $0.87 \pm  0.05$\\
 & $ 0.35$ & $0.87 \pm 0.09$ & $1.01 \pm 0.08$ & $0.94 \pm  0.09$ & $1.15 \pm  0.09$ & $0.98 \pm  0.08$ & $0.87 \pm  0.10$\\
 & $ 0.45$ & $0.91 \pm 0.04$ & $0.92 \pm 0.11$ & $1.00 \pm  0.13$ & $1.17 \pm  0.09$ & $1.01 \pm  0.08$ & $0.90 \pm  0.08$\\
 & $ 0.55$ & $0.96 \pm 0.09$ & $1.00 \pm 0.14$ & $0.99 \pm  0.11$ & $1.05 \pm  0.11$ & $1.04 \pm  0.09$ & $0.94 \pm  0.06$\\
 & $ 0.65$ & $1.06 \pm 0.10$ & $1.04 \pm 0.14$ & $1.04 \pm  0.12$ & $1.00 \pm  0.11$ & $1.02 \pm  0.06$ & $1.10 \pm  0.10$\\
 & $ 0.75$ & $1.12 \pm 0.12$ & $1.06 \pm 0.18$ & $0.96 \pm  0.15$ & $1.00 \pm  0.10$ & $1.08 \pm  0.08$ & $1.12 \pm  0.13$\\
 & $ 0.85$ & $1.34 \pm 0.19$ & $0.98 \pm 0.36$ & $0.88 \pm  0.15$ & $0.89 \pm  0.10$ & $1.02 \pm  0.11$ & $1.18 \pm  0.13$\\
 & $ 0.95$ & $1.56 \pm 0.22$ & $1.02 \pm 0.42$ & $1.08 \pm  0.20$ & $0.90 \pm  0.11$ & $0.99 \pm  0.25$ & $1.31 \pm  0.06$\\  
    \noalign{\smallskip}\hline\noalign{\smallskip}
$\varepsilon = 173\;\mathrm{MeV}$ 
 & $-0.95$ &                 & $2.46 \pm  0.42$ & $2.86 \pm  0.37$ & $3.14 \pm  0.56$ & $2.41 \pm  0.57$ & $2.47 \pm  0.37$\\
 & $-0.85$ &                 & $2.43 \pm  0.45$ & $2.28 \pm  0.47$ & $2.14 \pm  0.44$ & $2.51 \pm  0.33$ & $2.45 \pm  0.36$\\
 & $-0.75$ &                 & $1.89 \pm  0.43$ & $1.75 \pm  0.38$ & $2.26 \pm  0.65$ & $2.47 \pm  0.33$ & $2.25 \pm  0.32$\\
 & $-0.65$ &                 & $2.53 \pm  0.41$ & $2.13 \pm  0.44$ & $1.66 \pm  0.35$ & $2.07 \pm  0.29$ & $1.99 \pm  0.36$\\
 & $-0.55$ &                 & $2.33 \pm  0.38$ & $2.25 \pm  0.51$ & $2.15 \pm  0.38$ & $2.21 \pm  0.32$ & $2.13 \pm  0.31$\\
 & $-0.45$ &                 & $1.89 \pm  0.35$ & $2.04 \pm  0.39$ & $1.86 \pm  0.54$ & $2.25 \pm  0.32$ & $2.67 \pm  0.33$\\
 & $-0.35$ &                 & $2.08 \pm  0.36$ & $2.30 \pm  0.43$ & $2.49 \pm  0.42$ & $2.17 \pm  0.23$ & $2.10 \pm  0.34$\\
 & $-0.25$ &                 & $2.11 \pm  0.34$ & $2.10 \pm  0.45$ & $1.50 \pm  0.51$ & $1.56 \pm  0.36$ & $1.80 \pm  0.24$\\
 & $-0.15$ &                 & $1.98 \pm  0.41$ & $2.05 \pm  0.41$ & $2.20 \pm  0.37$ & $1.83 \pm  0.32$ & $1.72 \pm  0.27$\\
 & $-0.05$ &                 & $1.76 \pm  0.37$ & $1.61 \pm  0.46$ & $2.25 \pm  0.38$ & $2.13 \pm  0.39$ & $2.12 \pm  0.27$\\
 & $ 0.05$ & $1.80 \pm 0.53$ & $2.16 \pm  0.35$ & $1.61 \pm  0.46$ & $2.12 \pm  0.37$ & $2.13 \pm  0.39$ & $1.30 \pm  0.24$\\
 & $ 0.15$ & $2.37 \pm 0.49$ & $2.42 \pm  0.49$ & $2.05 \pm  0.41$ & $2.21 \pm  0.37$ & $1.83 \pm  0.32$ & $1.81 \pm  0.31$\\
 & $ 0.25$ & $2.28 \pm 0.39$ & $2.23 \pm  0.38$ & $2.10 \pm  0.45$ & $2.31 \pm  0.39$ & $1.56 \pm  0.36$ & $1.64 \pm  0.31$\\
 & $ 0.35$ & $1.96 \pm 0.31$ & $2.90 \pm  0.51$ & $2.30 \pm  0.43$ & $2.85 \pm  0.41$ & $2.17 \pm  0.23$ & $1.78 \pm  0.24$\\
 & $ 0.45$ & $2.41 \pm 0.40$ & $1.78 \pm  0.51$ & $2.04 \pm  0.39$ & $2.34 \pm  0.48$ & $2.25 \pm  0.32$ & $2.08 \pm  0.40$\\
 & $ 0.55$ & $1.87 \pm 0.29$ & $1.92 \pm  0.50$ & $2.25 \pm  0.51$ & $1.77 \pm  0.44$ & $2.21 \pm  0.32$ & $1.89 \pm  0.29$\\
 & $ 0.65$ & $2.24 \pm 0.35$ & $2.00 \pm  0.66$ & $2.13 \pm  0.44$ & $1.82 \pm  0.43$ & $2.07 \pm  0.29$ & $2.37 \pm  0.30$\\
 & $ 0.75$ & $2.27 \pm 0.39$ & $3.51 \pm  0.81$ & $1.75 \pm  0.38$ & $2.14 \pm  0.40$ & $2.47 \pm  0.33$ & $2.00 \pm  0.38$\\
 & $ 0.85$ & $3.04 \pm 0.36$ & $2.47 \pm  0.78$ & $2.28 \pm  0.47$ & $2.27 \pm  0.43$ & $2.51 \pm  0.33$ & $2.37 \pm  0.38$\\
 & $ 0.95$ & $4.04 \pm 0.41$ & $2.06 \pm  1.24$ & $2.86 \pm  0.37$ & $1.49 \pm  0.43$ & $2.41 \pm  0.57$ & $3.29 \pm  0.33$\\
    
    \noalign{\smallskip}\hline
  \end{tabular}                 
\end{table*}                       
\clearpage

\begin{table}[b]
  \caption{Differential cross sections in $\rm\mu b/sr$ for the orientation of the $\omega$ decay plane
   at the different excess energies. The central value of a cosine interval of 0.2 is listed.}
  \label{tab:diffdp}
  \begin{tabular}{@{}rlll}      
    \hline\noalign{\smallskip}                            
    $\cos\gamma$ & $\varepsilon = 92\;\mathrm{MeV}$ & $\varepsilon = 128\;\mathrm{MeV}$ & $\varepsilon = 173\;\mathrm{MeV}$\\
    \noalign{\smallskip}\hline\noalign{\smallskip}
      $ 0.1$ & $0.58 \pm 0.45$ & $0.65 \pm 0.09$ & $2.21 \pm 0.57$\\
      $ 0.3$ & $0.35 \pm 0.17$ & $0.76 \pm 0.06$ & $1.78 \pm 0.48$\\
      $ 0.5$ & $0.87 \pm 0.18$ & $0.96 \pm 0.04$ & $2.13 \pm 0.28$\\
      $ 0.7$ & $0.70 \pm 0.15$ & $1.04 \pm 0.03$ & $2.21 \pm 0.45$\\
      $ 0.9$ & $0.94 \pm 0.15$ & $1.28 \pm 0.03$ & $2.85 \pm 0.33$\\
    \noalign{\smallskip}\hline
  \end{tabular}                 
\end{table}                       

\begin{table}[b]
  \caption{Differential cross sections in $\rm\mu b/sr$ for the invariant mass
           of the $\omega$p-system, statistical errors are given only. The central mass 
           of an interval of 10 $\rm MeV/c^2$ is listed.}
  \label{tab:diffMwp}
  \begin{tabular}{@{}rlll}      
    \hline\noalign{\smallskip}                            
    $M_{\omega p}$  $(\rm MeV/c^2)$ & $\varepsilon = 92\;\mathrm{MeV}$ & $\varepsilon = 128\;\mathrm{MeV}$ & $\varepsilon = 173\;\mathrm{MeV}$\\
    \noalign{\smallskip}\hline\noalign{\smallskip}
 $1715$ & $0.16 \pm 0.07$ & $0.16 \pm 0.01$ & $0.30 \pm 0.07$\\
 $1725$ & $0.45 \pm 0.08$ & $0.40 \pm 0.01$ & $0.54 \pm 0.10$\\
 $1735$ & $0.91 \pm 0.11$ & $0.78 \pm 0.02$ & $1.21 \pm 0.17$\\
 $1745$ & $1.18 \pm 0.12$ & $1.20 \pm 0.02$ & $1.58 \pm 0.24$\\
 $1755$ & $1.28 \pm 0.12$ & $1.19 \pm 0.02$ & $1.34 \pm 0.22$\\
 $1765$ & $1.17 \pm 0.13$ & $1.31 \pm 0.02$ & $1.97 \pm 0.24$\\
 $1775$ & $1.26 \pm 0.13$ & $1.27 \pm 0.02$ & $1.65 \pm 0.24$\\
 $1785$ & $1.37 \pm 0.13$ & $1.27 \pm 0.02$ & $2.52 \pm 0.25$\\
 $1795$ & $1.24 \pm 0.11$ & $1.28 \pm 0.02$ & $2.07 \pm 0.24$\\
 $1805$ & $0.78 \pm 0.10$ & $1.13 \pm 0.02$ & $1.89 \pm 0.23$\\
 $1815$ &                 & $1.25 \pm 0.02$ & $2.08 \pm 0.23$\\
 $1825$ &                 & $0.99 \pm 0.02$ & $1.67 \pm 0.23$\\
 $1835$ &                 & $0.64 \pm 0.02$ & $2.26 \pm 0.23$\\
 $1845$ &                 & $0.42 \pm 0.02$ & $2.11 \pm 0.22$\\
 $1855$ &                 &                 & $1.74 \pm 0.20$\\ 
 $1865$ &                 &                 & $1.81 \pm 0.20$\\
 $1875$ &                 &                 & $1.49 \pm 0.18$\\
 $1885$ &                 &                 & $0.86 \pm 0.15$\\
                                                          
    \noalign{\smallskip}\hline
  \end{tabular}                 
\end{table}                       

\begin{table}[b]
  \caption{Differential cross sections in $\rm\mu b/sr$ for the invariant mass 
           of the pp-system, statistical errors are given only. The central mass 
           of an interval of 10 $\rm MeV/c^2$ is listed.}
  \label{tab:diffMpp}
  \begin{tabular}{@{}rlll}      
    \hline\noalign{\smallskip}                            
    $M_{pp}$  $(\rm MeV/c^2)$ & $\varepsilon = 92\;\mathrm{MeV}$ & $\varepsilon = 128\;\mathrm{MeV}$ & $\varepsilon = 173\;\mathrm{MeV}$\\
    \noalign{\smallskip}\hline\noalign{\smallskip}

$1882$ & $0.95 \pm 0.15$ & $0.69 \pm 0.02$ & $1.27 \pm 0.23$\\
$1892$ & $1.05 \pm 0.15$ & $0.77 \pm 0.02$ & $1.29 \pm 0.24$\\
$1902$ & $1.29 \pm 0.17$ & $1.10 \pm 0.03$ & $0.83 \pm 0.26$\\
$1912$ & $1.20 \pm 0.18$ & $1.03 \pm 0.03$ & $1.07 \pm 0.25$\\
$1922$ & $1.32 \pm 0.18$ & $1.24 \pm 0.03$ & $1.73 \pm 0.28$\\
$1932$ & $1.44 \pm 0.18$ & $1.59 \pm 0.03$ & $2.77 \pm 0.29$\\
$1942$ & $0.94 \pm 0.15$ & $1.41 \pm 0.03$ & $1.99 \pm 0.35$\\
$1952$ & $1.25 \pm 0.17$ & $1.34 \pm 0.03$ & $2.40 \pm 0.31$\\
$1962$ & $0.41 \pm 0.15$ & $1.30 \pm 0.03$ & $2.48 \pm 0.33$\\
$1972$ & $0.01 \pm 0.16$ & $1.11 \pm 0.03$ & $1.70 \pm 0.32$\\
$1982$ &                 & $0.97 \pm 0.03$ & $2.04 \pm 0.31$\\
$1992$ &                 & $0.44 \pm 0.02$ & $2.60 \pm 0.31$\\
$2002$ &                 & $0.23 \pm 0.01$ & $2.07 \pm 0.32$\\
$2012$ &                 &                 & $2.00 \pm 0.33$\\
$2022$ &                 &                 & $1.55 \pm 0.30$\\
$2032$ &                 &                 & $0.89 \pm 0.26$\\
$2042$ &                 &                 & $0.91 \pm 0.25$\\                         

    \noalign{\smallskip}\hline
  \end{tabular}                 
\end{table}                       

\end{document}